A SIMPLE EQUATION TO CALCULATE THE DIAMETERS OF BIOLOGICAL VESICLES.

Willy S. Bont

Division of Immunology and Biophysics,

The Netherlands Cancer Institute,

Plesmanlaan 121, 1066 CX Amsterdam

The Netherlands

wsbo@hetnet.nl

Running title: DISCRETENESS OF VESICLES

Summary

The remarkable preference of biomembranes, to constitute vesicles of certain discrete sizes is explained by using the following properties of phospholipids that are either well understood or at least documented.

A. By hexagonal close-packing their fatty acyl chains form a triangular lattice.

Their molecules:

B. Prefer to form linear arrays that occasionally make angles of 120º.

C. Form relatively large hexagons.

Based on these properties a model for monolayers is proposed and a simple equation derived for the calculation of diameters of vesicles. The diameters of vesicles of neurotransmitters and hormones determined by electron microscopy were compared with those obtained with the equation. Statistical analysis of this comparison revealed the model to give very significant results (p=.0002).

**Key words** monolayers-crystals-hexagons-cholesterol-vesicles-discreteness

## 1. Introduction.

The dimensions of vesicles may range from the small granular vesicles containing neurotransmitters and hormones to large structures like plasma membranes or nuclear membranes. The larger membrane structures are fragmented in the course of their isolation resulting in preparations of so-called in vitro vesicles with a broad distribution of sizes. These preparations were subjected to sedimentation analysis by analytical centrifugation; this analysis is based on differences in mass of sedimenting particles. Sedimentation analysis of preparations of biomembranes revealed that the mass (and therefore the surface area) of in vitro vesicles does not have a continuous distribution but has discrete values [7]. In the first publication only sedimentation coefficients were given; diameters were not yet calculated.

In the next paper [6] sedimentation coefficients were converted into mass. And since the mass of empty vesicles is proportional with their surface area, the latter quantity could be calculated from sedimentation analysis. Therefore in this paper size refers in the first place to surface area; diameter or volume can be calculated from the surface area using the equations for spheres.

The results from the sedimentation analyses were unexpected. Quantitative analysis of the data revealed that the surface areas of the vesicles constitute two geometric series with two adjacent terms differing by a factor 2. The two series can be represented by equations

$$A(n) = A(1)2^{(n-1)} \text{ (=series 1)} \qquad \text{(equation 1)}$$

$$A'(n) = 3A(1)2^{(n-1)} \text{ (=series 2)} \qquad \text{(equation 2)}$$

where $A(1)$ = surface area of smallest vesicles

$A(n)$ = surface area of the nth term of series 1

$A'(n)$ = surface area of the nth term of series 2

It could be demonstrated that the size of all vesicles found in vivo (not only for neurotransmitters and hormones but also of viruses, chlamydiae, nuclei and even whole cells) fit in the very same two geometric series observed for fragmented membranes. All these aspects and the experimental procedure have been reviewed previously [5].

The purpose of this publication is to explain these peculiar results and to derive one single, relatively simple, equation to calculate quantitatively the diameters of virtually all in vivo vesicles. This investigation is based on the assumption that the size of in vivo vesicles is determined by properties of the various components and their interactions.

## 2. Properties common to membranes.

Since it is assumed that the size of vesicles is primarily determined by phospholipids, this section deals with the properties of phospholipids that are important for the mechanism proposed in this paper.

It is an important characteristic of all membrane vesicles that their phospholipids form bilayers [12]. Lipids of either monolayer show at least at some temperatures and lipid compositions the following three aspects:

In the first place the fatty acyl chains in preparations of lecithins can form a two-dimensional hexagonal lattice [32]. In this paper it is assumed that by hexagonal close packing (Fig. 1) the latter lattice exists in virtually all in vivo vesicles.

Secondly, bands of the linear arrays formed by molecules of phospholipids can occasionally change direction by making an angle of 120º [35]. In this paper it is assumed that angles of 120º are formed by linear arrays of phospholipids in virtually all vesicles (Fig. 1C).

Thirdly, the electric dipoles attached to headgroups of the PL- molecules are coupled and form zig-zag ribbons [16].

It is assumed that due to the versatility of the headgroups, their dipoles do not interfere with, but rather adapt themselves to, the direction of the glycerol moieties. The latter is only determined by the direction of the arrays of the chains that is either linear or forms angles of 120º.

### 3. An explanation for the widespread occurrence of hexagons on membranes.

Hexagonal patterns have been observed by microscopy on biomembranes from sources as different as thrombocytes [27] and the endoplasmic reticulum from trophosphongial cells of kangaroo rat placenta [33]. Hexagons in the form of a honeycomb were also observed on membranes of pure phospholipids with cholesterol [31].

In the following it is demonstrated that given a hexagonal close-packing of the chains, relatively large honeycomb-like patterns must be formed in all phospholipid monolayers. This occurs if the two following very important conditions are fulfilled simultaneously. First, hydrophobic interactions must take place to the largest extent and therefore vacancies must be minimized. Secondly, maximal dipole interactions between the headgroups should favor the formation of closed linear arrays of molecules and therefore two open ends (one with a negative charge and the other with a positive charge) of a linear array of PL should be avoided. A set of linear parallel arrays of PL on a vesicle (like the lines of latitude on a globe) fulfills these two conditions. If angles of 120º are introduced the second condition can only be realized if the PL form a closed line that forms a hexagon. A hexagonal structure is here defined as a hexagon, filled with smaller hexagons formed by nesting. This results in a surface competely covered by hexagonal structures. Although on statistical grounds all kinds of hexagons can be formed the formation of regular hexagons must be preferred since with this type, by nesting, only one single vacancy in the middle of each hexagonal structure is formed (Fig. 1D); all other hexagons will lead to structures with open ends and/or more than one vacancy (Fig. 1E). There should be a preference for identical, regular hexagons (i.e. a honeycomb-like structure here called a honeycomb) to fill the surface of membranes to the greatest extent; otherwise space between hexagons will give rise to linear arrays with open ends and/or too many vacancies. Such a preferential formation of honeycombs has farreaching consequences.

Before calculations can be made, first the hexagonally close-packing of Fig 1 is represented by a triangular lattice as shown in Fig. 2.

A few relations are derived between the size of a monolayer and the number of different honeycombs that can cover it to the greatest extent. The starting point for the calculations is a flat monolayer. The following two assumptions are made: in the first place it is assumed that the complete monolayer is covered by a triangular lattice and secondly that the total monolayer can be represented by only a small part. In other words from a careful analysis of only a small part, many properties of a complete monolayer can be derived. This small part must be representative for, and contain all essential and relevant features of, the monolayer. Two aspects of the representative fragment are important for the analysis: its form and its size. The form is dictated by the triangular lattice and must be a regular triangle as shown in Fig.2C; a round or rectangular fragment cannot be reconciled with a triangular lattice. The second aspect is that the number of different hexagons that can be constructed in a regular triangle depends on its size, as shown in the following.

The size of a triangle, T, is given by $n_T$, the number of lattice-constants of one of the sides (in Fig. 2C, $n_T=18$). The size of the largest regular hexagon, H, that fits in T is represented by $n_H$, the number of lattice-constants of one of its sides (in Fig. 2C, $n_H=6$).

From geometrical considerations it follows that

$$n_T = 3n_H \quad \text{(equation 3)}$$

Instead of by one large hexagon, T can also be covered by honeycombs ,hon, where $n_{hon}$ is the number of lattice-constants of one its hexagons (in Fig. 2D, $n_{hon}=3$). It is a configuration with T covered by hexagons to the greatest extent while the remainder is filled with linear arrays. Later these honeycombs are discussed in more detail.

For $n_{hon}$ it can be shown that

$$n_T = (2(f(n)-1)+f(n)+2)n_{hon} = 3f(n)n_{hon} \quad \text{(equation 4)}$$

where f(n) is the number of hexagons of the honeycomb along a side of T. Equation (4) can be checked with Fig. 2D where f(n)=2. From equations (3) and (4) it follows that

$$n_H = f(n)n_{hon} \quad \text{(equation 5)}$$

It appears that the $n_{hon}$-values are factors of $n_H$. For another analysis of equation (3) see section 9.

4. Theory for the existence of two types of crystal lattice.

Only the size of the headgroup is important for this discussion and therefore it is restricted to only two of the many different diglycerides. As a representative of the diglycerides with a small headgroup phosphatidylethanolamine (PE) is given, while phosphatidylcholine (PC) represents diglycerides with a large headgroup. It is assumed that for in vivo vesicles the fatty acyl chains of both types of diglycerides form by close-packing the same triangular lattice. The influence of the difference in size between PE and PC is discussed in literature [16].

The lattice discussed so far holds for PE and is designated as a normal lattice. It is now assumed that the PC molecules are oriented in a direction perpendicular to the one of the PE in order to accommodate their large choline headgroup (Fig 1B.) Also in this so-called perpendicular lattice the molecules can be arranged in three directions.

In Fig. 3 a theoretical lattice for PC is derived from the perpendicular lattice of Fig.1B. The only difference between a normal and a centered lattice that is important for our calculations is, that the surface area of a centered unit-triangle is 3 times the surface area of a normal unit-triangle. In summary: the same hexagonal close-packing could in theory lead to 2 different types of lattice.

## 5. Experimental evidence for two types of crystal lattice.

Whether a normal or a centered lattice is formed depends on the three following factors.

### a) The influence of the headgroup of PL.

For a discussion of the influence of the headgroup on molecular packing of phosphatidylethanolamine (PE) and phosphatidylcholine (PC) the reader is referred to the literature [16]. Only aspects of packing that are relevant for this paper are mentioned.

In PE the headgroup has about the same cross-section as the sum of the cross-sections of the two chains. Therefore PE forms a normal lattice (see Fig. 1A) with its chains simultaneously perpendicular to the surface and with a hexagonal close-packing. Because the headgroup of PC is relatively large, their fatty acyl chains could be tilted and then PC cannot form a normal lattice.

### b) The influence of the molecular structure of diglycerides.

The molecular structure was determined by X-ray diffraction of crystals of PE [18] and of PC under conditions where the chains of the latter are perpendicular to the surface [29]. In both cases the same "tuning-fork" configuration was observed, with chain 1 perpendicular to the surface while the initial part of chain 2 is parallel to the surface but bends of at the second carbon atom to become parallel to chain 1, as shown in Fig. 4.
Interactions with the C=O group of chain 2 take place easily due to its accessibility [29].

### c) The influence of molecules with one single hydrocarbon chain.

The reaction of free fatty acids with PC was described in literature. Mabrey and Sturtevant [24] observed by high-sensitivity differential scanning calorimetry (DSC) that the transition temperature of dipalmitoyl glycerophosphocholine (DPPC) is raised from 41.4 to 61.5° C by the addition of palmitic acid (PA) at a mole fraction of 0.67. The transition temperature of pure DPPE, the ethanolamine derivate, is 63.8° C. The authors drew the attention to the close resemblance between the 2:1 complex of PA:DPPC and pure DPPE. In addition to almost the same transition temperature the transition curves of both materials have also the same assymetric form. The authors postulated the hypothetical component 1/2(DPPC+2PA) to explain their results.

Another molecule with one single hydrocarbon tail is cholesterol (Chol). Since Chol is so important its interaction with PC is discussed here in more detail.

Above it was noticed that the carbonyl-group of chain 2 of PC reacts easily. For the interaction with Chol this was confirmed by infrared [9] and Raman [2] spectroscopy.

The results by Oldfield et al. [28] obtained by nuclear magnetic resonance studies showed that the "tuning-fork" model (see above) also holds in the lecithin-cholesterol system. Blume and

Griffin [3] regard Chol as a "spacer" that disrupts the packing of the chains in PE, leading to a disruption of the packing of the headgroups. As to PC, Presti et al. [30] regard Chol as a "filler" which is associated with a change in chain tilt. Hui and He [19] stated that by the addition of Chol to PC the tilting is abolished and the chains become oriented almost perpendicular to the surface of the bilayer. Also the many phases that occur in membranes have been studied extensively and especially the domains observed in mixtures of PC and Chol are here of interest. Dark-field electron microscopy showed ribbon-like structures in mixed DPPC-Chol bilayers, the ribbons being less than 100 nm wide [20]. Furthermore Chol-rich domains exist of linear arrays of molecules in which one row of PL runs parallel to each row of 1:1 sterol-lipid complex. This would reduce to rows of complex alone as the Chol content is increased above a mole fraction $X_{Chol}=1/3$ [30]. Especially the 1:2 complex measured by Hinz and Sturtevant [17] using DSC must be mentioned.

The existence of a centered lattice can be deduced from the data cited above from the literature and especially from the work by Mabrey and Sturtevant [24] and Hinz and Sturtevant [17] as is now explained.

The results obtained by Mabrey and Sturtevant [24] with DPPC and PA can be explained as follows. The close resemblance between the 2:1 complex of PA:PC and pure DPPE can be understood if it is assumed that the properties at the transition temperature are in the first place determined by the packing of the chains. If this assumption is correct both in DPPE and in the hypothetical component an identical hexagonal close-packing of the chains must exist with the chains perpendicular to the surface. Apparently by DSC, in DPPE the same hexagonal close-packing of the chains is detected as in DPPC after the addition of PA. The difference is, that with the same close-packing, in pure DPPE a normal and in the mixture PA/DPPC=2 a centered lattice is formed.

The hypothetical component 1/2(DPPC+2PA) (cf. [24]) has not to be postulated as demonstrated by the following calculations. In a centered triangular lattice each unit-triangle has one chain in a center or n(cen)=1. Because every corner of a unit-triangle belongs to 6 triangles (see Figs 2 and 3) it represents 1/6 chain and one unit-triangle has 1/6*3=1/2 chains or n(cor)=1/2. Therefore the following holds:

$$n(cen)/n(cor)=2$$

It is more relevant to express this relation in molecular terms and not only in chains. Then it must be realized that a chain in a center belongs to one molecule of free fatty acid and the chains in the corners to 1/2 molecule of diglyceride. With the aid of equation (6) it can now be written for one unit-triangle:

$$\text{(molecule in center)/(molecules in corner)}=4 \quad \text{(equation 6)}$$

Because the total lattice can be represented by one of its unit-triangles, this also holds for the total lattice. Therefore for the whole lattice: (molecules in centers)/(molecules in corners)=4.
A centered lattice formed by PA and DPPC is shown in Fig. 5.

If all lattice points are occupied (Fig. 5A), from equation (6) it follows that the ratio PA/DPPC=4, corresponding with a mole fraction $X_{PA}=0.8$. However the hypothetical component postulated by Mabrey and Sturtevant [24] is formed at a mole fraction $X_{PA}=0.67$. This can be explained if it is assumed that not all centers of the lattice have to be occupied by PA. If the centers between two arrays of PC are alternately occupied by PA or empty, the ratio PA/DPPC=2 (Fig. 5B), corresponds with the mole fraction $X_{PA}=0.67$ observed by the authors.

Fig. 5B visualizes the stoichiometry of the interaction between PA and DPPC; of course this hypothetical configuration with so many vacancies will never be realized. A tentative mechanism for the interaction is that two consecutive processes take place. First the incorporation of molecules of PA into centers takes place followed by a transition of molecules of DPPC from the tilted form to the perpendicular configuration (see Fig. 1B). Only the latter process is monitored by DSC. The first process, the incorporation of PA, continues untill all DPPC is in the perpendicular form (see Fig. 5C). Further incorporation of PA is not monitored by DSC because in the interval between Fig. 5C and Fig. 5A the same hexagonally close-packed lattice exists. One has to keep in mind that DPPC with only PA is an artificial system in which not all centers of the lattice have to be occupied. It can be imagined that in vivo in addition to free fatty acids also other single-chained molecules are present to fill the potentially available vacant centers.

In a centered lattice more space is available for the large headgroup of PC. In a perpendicular lattice the chains of PC are normal to the surface and the molecules can form hexagons.

In order to construct a triangular centered lattice, two conditions must be fulfilled simultaneously: a gap between the two chains of one PC-molecule has to be enforced and close-packing must exist. Chol can fulfill both conditions when it forms a 1:1 complex with a PC-molecule, as illustrated in Fig. 6.

In Fig. 7 the formation of a centered lattice by 1:1 complex is shown.

Apparently Chol is used both as a "spacer" to keep the two chains in one PL molecule apart and as a "filler" to occupy the centers of the triangular lattice and to fix their hydrocarbon chains in a row to their lattice points, creating hexagonal close-packing. In Fig. 7A a centered lattice with linear arrays of 1:1 complexes is formed at $X_{Chol}$=0.5. From the work of Hinz and Sturtevant [17] it must however be concluded that already at $X_{Chol}$=.33 all lecithin molecules are forced into another type of triangular lattice. This corresponds to a ratio Chol/PC =1/2 or to a situation where two linear arrays of 1:1 complex alternate with two arrays of pure lecithin as shown in Fig. 7B. Due to the empty arrays (reserved for the centers), the lattice has to condense as shown in Fig. 7C. One has to keep in mind that DPPC with only Chol is an artificial system in which not all centers of the lattice have to be occupied; a mixture of the pure perpendicular lattice of Fig.1B and the centered lattice of Fig. 3 is possible. It can be imagined that in vivo in addition to Chol also other single-chained molecules like monoglycerides, lipoproteins or free fatty acids are present to fill the vacant centers.

For the formation of a centered lattice the mole fraction of PA required is twice that of Chol (0.67 against 0.33). This has to be related to the fact that Chol can perform two functions (it is both a "filler" and a "spacer") and PA only one (it only fills the centers); more PA is required (i.e. more centers have to be occupied) to keep the two fatty acyl chains of every molecule of DPPC separated.

6. An odd number cannot be a factor of $n_H$.

It is assumed that like for all other systems, also for membranes it holds that the free energy must be minimal and the entropy maximal. In other words they strive for the largest number of configurations (including the largest number of hexagons!) that is allowed by the energy required for configurations. In the case of the triangular fragments representative for the lipid monolayers of biomembranes as discussed above it means that they must have the largest number of different hexagons (honeycombs) i.e. $n_H$-values with the largest number of factors.

Fig. 8 illustrates why odd factors cannot be reconciled with stable monolayers.

Because two vicinal lattice points must correspond with one PL-molecule, every hexagon of even-honeycombs must have one vacancy whereas odd-honeycombs have none. By the presence of vacancies even-honeycombs are stabilized as will now be explained.

The formation of two open ends of an array must be avoided; they represent a free positive charge at one end and a free negative charge at the other end, prevent maximal interaction of the dipoles, and thus increase the energy. Of course open ends are formed at both sides of vacancies where dipole interactions are interrupted. But at vacancies the distance between two charges is very small. Furthermore vacancies diffuse over the whole surface and the charges are not fixed. Because they cannot be pinpointed to particular lattice points this kind of open ends only formally exist but are not important for this discussion. Only the open ends that are separated by a large distance and that are fixed at sites where two arrays meet cause instabilities, as illustrated in Fig. 9 for the hexagonal arrays of odd honeycombs.

The open ends in the corners of hexagons of odd-honeycombs cannot be avoided. In even-honeycombs these instabilities do not occur due to their vacancies, as is now explained.

With only 3 hexagons both in Fig. 8A and in 8B it was demonstrated that even-honeycombs have one vacancy in every hexagon. Now it is shown with the aid of Fig. 10, that vacancies exist in all the hexagons of even-honeycombs.

Vacancies can have various effects. For instance they enable phospholipid molecules to perform various movements like translocations and rotations that result in their diffusion [23] and this in turn causes the molecules to be more loosely packed. In other words a vacancy increases the surface area of a monolayer more than its own small contribution.

The distribution of vacancies shown in Fig. 10 is only one of the many that are possible. Due to diffusion the number of vacancies is not the same in every hexagon but at the average there is exactly one vacancy per hexagon. In practice some hexagons have for instance 2 vacancies while others have none. The latter hexagons are isolated structures coupled by arrays that can contain a variable number of vacancies. Therefore, in even-honeycombs the direction of the dipoles has not the effect it has in odd-honeycombs. Despite the relatively small number of vacancies in even honeycombs both the number of different stable configurations and the entropy are increased enormously. The most realistic picture for an even-honeycomb is a very dynamic one due to the presence of a small number of vacancies, evenly distributed over the whole honeycomb.

An odd-honeycomb with $n_{hon}=3$ is the only one that cannot lead to a situation shown in Fig. 9. It cannot be formed because due to the large number of vacancies of the whole membrane resulting from the vacancies in the middle of every hexagonal structure, and due to the formation of a large number of angles of respectively 60° and 120°, the energy required for its formation is too high. Because a honeycomb with $n_{hon}=3$ cannot lead to instable configurations, $n_H$ can have a factor 3.

In summary: when an integer is divided by its odd factors it becomes a term of the geometric series $2^n$. By multiplying the terms of this geometric series (with n even factors) with 3, a second geometric series $3*2^n$ is formed while the number of even factors is doubled.

All these considerations, confining the $n_H$-values to terms of the two geometric series $2^n$ and $3*2^n$, result in the following two combined series of $n_H$-values for triangular monolayers:

$$1,2,3,4,6,8,12,16,24,32,......,2^n,3*2^{n-1},2^{n+1} \text{ (series (a))}.$$

7. A comparison of surface areas of triangular monolayers and vesicles.

Series (a) was derived for flat, two-dimensional triangular monolayers. Real biomembranes like vesicles or cells are however formed by two monolayers that form a three-dimensional body. The results obtained for triangular monolayers may be extrapolated to real biomembranes, only if each of their two, three-dimensional monolayers are composed exclusively of two-dimensional triangles that fit in series (a). Calculations must show that this theory is correct and that series (a) not only holds for triangular monolayers but also for real biomembranes.

Since in practice it are the surface areas of vesicles that fit in two geometric series the discussion in this section is restricted to surface areas. Diameters or volumes must be derived with the equations for spheres.

The surface area of a certain triangle, T, expressed as the number of unit-triangles (see Fig. 2) is represented by S(T) and is given by the sum of the arithmetic series with terms (2n-1) where n is from n=1 to n=$n_T$ or

$$S(T)=\Sigma (2n-1)=n_T*(1+(2n_T-1))/2=n_T^2$$

Since $n_T=3n_H$ (see equation (3)) the surface areas are also given by

$$S(T)= 9(n_H)^2 \text{ (equation 7)}$$

It is evident that the surface area of a triangle, T, with a centered lattice, is 3 times the surface area of a triangle with a normal lattice if both have the same number of unit-triangles.

Furthermore the $n_H$-values of stable triangles form the series (a) (see above) and therefore their surface areas form the series:

$$1,4,9,16,36,64,144,256.......9*2^{2(n-1)},2^{2(n+1)} \text{ (series (b))}$$

and for triangles with a centered lattice::

$$3,12,27,48,108,192........,27*2^{2(n-1)},3*2^{2(n+1)} \text{ (series (c))}$$

Now the results derived for the theoretical surface areas of flat triangular monolayers, series (b) and series(c), are compared with those experimentally obtained for the surface areas of real vesicles.

In practice the surface areas of vesicles were determined both by analytical centrifugation via the mass of in vitro vesicles in preparations of biomembranes and by electron microscopy via the diameter of in vivo vesicles for neurotransmitters and hormones, and it was shown [4] that both types of vesicles fit in the same two geometric series

$$1,2,4,8,16,32,64,128,256.......2^{2n} \text{ (series (d))}$$

and a series

$$3,6,12,24,48,96………3*2^{2n} \text{ (series (e))}$$

Series (d) and series (e) are represented by equations (1) and (2).

Between half the values of series (b) for flat triangular monolayers with a normal lattice and for vesicles of series (d), there is a difference of 12.5%; the theoretical average difference between all terms of the two series is 6%. In practice, the error in the mass of a vesicle determined by analytical centrifugation is about 6% [34]. Therefore, since it is difficult to differentiate between series (b) and series (d) the former can be approximated by equation (1) and corresponds to what was previously called series 1. The same holds for series (c) and series (e); series (c) can be approximated by equation (2). It must be concluded that the surface areas of both theoretical triangular monolayers and real vesicles can be represented by the same two geometric series. Is this result a coincidence or is there a connection between the discreteness of vesicles and the discreteness of triangular monolayers? A comparison between flat triangular monolayers and spherical bilayers requires further analysis.

Until now the theory was restricted to hypothetical, triangular flat monolayers. In practice, membranes formed by phosphodiglycerids are closed, often spherical bodies (vesicles) confined by bilayers and not by monolayers. But a vesicle is just two back-to-back spherical monolayers with negligible interaction between the two [26]. Therefore for our calculations a suspension of bilayers is considered to be a suspension of spherical monolayers.

For a comparison between flat monolayers and spherical membranes the following aspects are important.

First of all it is postulated that spherical monolayers are composed of triangular fragments (as discussed above) and that hexagons must be formed to the greatest extent on both systems.

Secondly the surface areas of flat triangular monolayers have to be expressed in the same units as the spherical monolayers i.e. in $nm^2$ and not in the number of unit-triangles. To express unit-triangles in $nm^2$ it must be realized that every corner of a unit-triangle represents a fatty acyl chain. This chain must be divided by 6 unit-triangles, and the 3 corners of a unit-triangle therefore represent 3/6= 1/2 fatty acyl chain. The cross-sectional area of a chain is 0,25 $nm^2$ [16] and therefore the surface area of every unit-triangle of a normal lattice = 0.125$nm^2$ .The real surface areas of triangular monolayers are respectively $0.125*n_T^2 = 0.125*9*n_H^2 = 1.125*n_H^2$ $nm^2$ for a normal lattice (series (b)) and $3.375*n_H^2$ $nm^2$ for a centered lattice (series (c)).

The third problem is the determination of the effective diameter of spherical vesicles. An important aspect of this question is whether this diameter is closer to the inner layer then to the outer layer. Since the theory must be applied to both layers an important condition is that both layers must have the same number of molecules of diglycerides. Keeping the latter condition in mind (see also Discussion), it is assumed that the size of vesicles is determined by their inner monolayers. Because the inner monolayers are the smallest, their lipids are more tightly packed and therefore they have the smallest number of vacancies, corresponding with the largest possible hexagons. The larger outer monolayers can adapt their surface areas in several ways, as discussed in the Discussion

The, for our calculations important, relevant surface area of a vesicle is determined by $D^2$ where D is the effective diameter of the inner monolayer. D is located closer to the glycerol moiety than to the middle of the bilayer because the hydrophobic fatty acyl chains in membranes exhibit rapid isotropic motion in the middle of the bilayer whereas in the same monolayer the chain motion near the glycerol moiety can approach that of the solid, crystalline hydrocarbons [25] on which the model is based. The middle of a bilayer is 5nm from the outer diameter and D= D(V)-12 nm where D(V) is the outer diameter of the vesicles that is measured in practice (see Fig. 11).

Now a comparison between the surface areas of respectively vesicles and their flat triangular components is possible as is illustrated in Table 1.

The following vesicles are used: synaptic vesicles from cat [13]; synaptic vesicles from rat [8]; hormone vesicles from rat [11] and human hormone vesicles [14].

From Table 1 it must be concluded that the inner layers of in vivo vesicles consist of 20 identical triangular building blocks when their $n_H$ -values are terms of series (a).

## 8. A model for vesicles.

Since the diameters of vesicles are determined by the size of their inner layers and the latter are composed of 20 identical triangles, it is tempting to speculate that the model that best describes an in vivo vesicle is an icosahedron, one of the 5 regular polyhedrons. All problems related to the implication of this model cannot be solved in this paper. But an aspect that seems to contradict the reality of the model is now discussed. The first most obvious objection is that vesicles are spherical; icosahedral vesicles have never been observed. Therefore it is required to investigate how much a spherical vesicle deviates from an icosahedron if both have the same surface area. In Fig. 12 a cross-section is presented across an icosahedron and the sphere with the same surface area.

In general the surface areas almost coincide and it will now be discussed that the small difference between the two cannot be used as an argument against the model.

It is assumed that hexagonal close-packing of the chains leading to a triangular crystal-lattice, not only holds for flat monolayers but also for both layers of vesicles. Furthermore the icosahedron model can only exist when all essential characteristics of the icosahedrons are preserved as good as possible in the spherical vesicles. The cross section in Fig. 12 shows that in some directions the icosahedron and the sphere intersect. Since it is required to preserve the triangular crystal lattice on the icosahedrons as good as possible, it is projected on the spheres. In our calculations with flat, triangular monolayers all unit-triangles were supposed to have the same size; in contrast the dimensions of unit-triangles projected on the spheres cannot be identical. In some directions where the distance to the sphere is larger the unit-triangles must be larger and in other places, especially in the directions of the 12 corners, they must be smaller. This requires further explanation.

First of all in the calculations for the conversions of unit triangles in $nm^2$ the cross-sectional area of a chain is 0.25 $nm^2$ [16]. This value is not a universal constant like many constants often used in physics. It is an average value for membranes in the liquid-crystalline state; in the condensed phase, at lower temperatures it ranges even from about 0.185 to 0.21 $nm^2$ [16]. Another reason for a spread in the cross-sectional area could be ascribed to the heterogeneity of the fatty acyl chains as present in biomembranes. Could the composition of the diglycerides and thus the cross-sectional area be slightly different at various sites of the vesicles? It must be concluded that on spherical vesicles the same triangular lattice can exist as on icosahedral vesicles albeit that the size of the unit-triangles is not constant. Since however the differences between icosahedrons and the corresponding spheres are so small the average size of the unit-triangles on both systems is identical.

Another justification for the use of the icosahedron model for spherical vesicles is given by the fact that membranes observed in vivo are liquid-crystals [16]. This implies that they have properties in common both with liquids and with crystals. It is well known that a small drop of liquid in a suspension forms a sphere; the driving force behind this phenomenon is the surface tension. Now it is postulated that on small "drops" of liquid-crystal (i.e. the icosahedral vesicles) the surface tension creates the spherical shape, keeping the triangular lattice intact.

In conclusion: icosahedrons can be models for spherical vesicles. In practice the advantage is that on one hand icosahedrons can be considered as spherical particles while on the other hand calculations have not to be performed with the intact vesicles but can for our purpose be restricted to one of the twenty identical triangles. This will be discussed in more detail in the following section.

### 9. Visualization of a three-dimensional icosahedral vesicle.

To make it more understandable that only few of the many possible configurations of phospholipids lead to stable vesicles, a three-dimensional icosahedron must be visualized in a two-dimensional representation. In Fig. 13 the 20 triangles of an icosahedron are given in a plane; with the aid of this figure a three-dimensional icosahedron can be constructed.

In Fig. 14 only 10 of the 20 triangles of an icosahedron are used.

With the same hexagonal close-packing of the fatty acyl chains and the same triangular lattice over the whole surface of the icosahedral membrane, the size of the hexagon is important as shown in Fig. 15.

Basically the argument is that the configuration in Fig. 14 corresponds to a more stable packing of lipids than the configuration of Fig. 15 because there are no open ends. If only one of the 10 triangles in Fig. 15 is placed in a different position (by rotation over an angle of 120º or 240º), the number of open ends is already changed.

In the triangle of Fig. 15A, in addition to a hexagon and the three linear arrays at the 3 corners, three extra arrays (with two angles of 120º and therefore non-linear!) are present. One of the latter three arrays is the source of open ends on the icosahedron.

If a monolayer can be represented by 20 icosahedral triangles with $n_T$-values without a factor 3 it is impossible to make a configuration without open ends. For instance with $n_T=14$ (figure not shown) $n_H=4$ is the largest regular hexagon that can be used. The situation is now comparable with the one shown in Fig. 15A except for the fact that every triangle has two, instead of three, extra non-linear arrays with two angles of 120°. These arrays are again a source of open ends and therefore in general all icosahedrons with $n_T$-values without a factor 3 must have a relatively high energy and cannot be stable.

Figs 14 and 15 make it understandable that only for $n_T$-values with a factor 3, configurations without open ends and therefore with low energy, can be constructed. Apparently equation (3) not only follows from purely geometrical considerations, but also for energetical reasons the relation $n_T = 3n_H$ must hold.

It can be concluded that a symmetrical arrangement of phospholipids can only be fulfilled for configurations like the one in Fig. 14. Only with the latter configurations calculations for the whole vesicle can be restricted to one triangle (see previous section).

### 10. Fusion and budding of vesicles.

During fusion two in vivo vesicles disappear and become one vesicle. The opposite, the appearance of new in vivo vesicles also occurs, for instance by budding of virus from the outer membranes of cells.

As to in vitro vesicles it is assumed that the same two processes, disappearance and appearance of vesicles occurring in vivo, also hold for in vitro vesicles. Now the two processes are designated respectively as fusion and disintegration.

In this paper in vitro vesicles are mentioned only occasionally; they were not and will not be used for the development of the model for monolayers and its application to in vivo vesicles. However the results obtained by sedimentation analysis gave the first indication of serial discreteness. Therefore in the following it is speculated how the formation of in vitro vesicles could result from both fusion and disintegration.

Biomembranes are fragmented into discrete-sized vesicles during their isolation and it is assumed that this process proceeds in steps. The first step, the rough, rapid mechanical disruption of large structures is not a property of membranes. It is difficult to imagine that this rough fragmentation in vitro of large membrane structures leads instantaneously to discrete-sized, stable fragments and is here supposed to result in vesicles with a continuous distribution of sizes. It is assumed that all unstable vesicles have open ends and form discrete-sized, stable ones [6,34] in a second step either by disintegration or by fusion.

As an example of vesicles that disintegrate, those with an odd factor are now discussed. The disintegration of these vesicles is advantageous for the following reasons. First of all they are unstable when an odd-honeycomb is realized (cf Fig. 9). Secondly an odd factor like for instance 17 contributes only one factor to the $n_H$-value while a possible product of disintegration with the smaller number $16=2^4$, adds 4 factors. Of course disintegration only takes place if the total free energy of all products of disintegration is lower than the free energy of the vesicle that disintegrates. Ultimately, thermodynamics determines whether or not a vesicle disintegrates.

Now disintegration is studied in more detail. It must be kept in mind that it is the surface area that disintegrates. Therefore the sum of the surface areas of the products of disintegration equals the surface area of the starting material. The surface area is proportional with $n_H^2$ and therefore also with the square of any of its factors.

This is demonstrated with an $n_H$-value with the factor 7.

Because $7^2 = (3*2)^2 + 3^2 + 2^2$ for $n_H=7*2^n$ the following relation holds for its products:

series with $n_H=7*2^n$ = series with $3*2^{n+1}$ + series with $3*2^n$ + series with $2^{n+1}$

It must be concluded that disintegration of an unstable vesicle with $n_H=7*2^n$ and therefore with $2n$ even factors results in three stabe vesicles; one of them with a larger number (viz. $2n+2$) of even factors.

Vesicles with a factor 5 disintegrate according to:

series with $n_H = 5*2^n$ = series with $n_H = 3*2^n$ + series with $n_H = 2^{n+2}$.

In the latter example a series with $2n$ even factors disintegrates into two series: one with the same number of $2n$ even factors plus one with $n+2$ even factors.

In general: disintegration of vesicles with an odd factor results in at least one vesicle with either the same number of even factors (when odd factor=5) or more even factors (when odd factor >5). Disintegration results in at least two vesicles and only takes place, as already mentioned above, if the total free energy of the system is lowered by disintegration.

Fusion or disintegration of vesicles with a factor 3 is not a probable alternative because formation of a honeycomb with $n_{hon}=3$ is very unlikely due to its relatively large energy (see section 6 and Fig. 9). A factor 3 must even be preferred because by its presence the number of even factors and therefore the number of stable configurations is doubled.

It is supposed that disintegration of an unstable vesicle takes place by a rearrangement of its molecules; no external forces are required. For fusion two unstable vesicles are required and it is assumed that the conditions favoring fusion are present on both vesicles. It is postulated that for fusion the presence of open ends on both vesicles is the decisive factor. The mutual attraction of the electric charges present on both vesicles, leading to the annihilation of open ends, could be the driving force behind fusion.

The processes of disintegration and fusion take place simultaneously; a certain unstable vesicle can either disintegrate or fuse with another unstable vesicle. The $n_H$-values of all stable fragments of in vitro vesicles formed after exhaustive disintegration and fusion belong to series (a).

## 11. Theoretical diameters of vesicles.

Since in literature diameters of vesicles are given and seldom their surface areas it is easier to use diameters though surface areas are more relevant. For a comparison between theory and practice, diameters can be used only when the particles are spherical and the diameters can easily be converted into surface areas.

In Table 1 it was shown that by the calculation of the experimental surface areas from their measured diameters it appears that vesicles can be represented by icosahedrons. But in the light of this result it is of course also possible to postulate that vesicles are icosahedrons and to calculate their theoretical diameters as shown in the following.

The surface area of a theoretical icosahedral monolayer expressed as N(tri), the number of unit-triangles (see also equation (7)), is given by:

$$N(tri) = 20\, n_T^2 = 20*(3n_H)^2 \qquad \text{(equation 8)}$$

Since each unit-triangle of a normal lattice contains 1/2 chain derived from its three corners:

$$\text{surface area of each unit-triangle} = C/2 \qquad \text{(equation 9)}$$

where C = cross-section of a fatty acyl chain in $nm^2$.

Multiplication of (8) with (9) gives for the surface area, A, in $nm^2$:

$$A = 10*(3n_H)^2*C \qquad \text{(equation 10)}$$

On the other hand for a sphere with the same surface area, A, the following relation holds:

$$A = \pi D^2 \qquad \text{(equation 11)}$$

Equating (10) with (11) gives:

$$\pi D^2 = (3n_H)^2 * 10 * C$$

$$\text{or } D = 3n_H * \sqrt{(10C/\pi)} \qquad \text{(equation 12)}$$

With (12) the effective diameters of all stable monolayers with a normal lattice can be calculated with C as the only experimental constant.

Also the diameters of monolayers in series (c) can now be calculated because a centered unit-triangle is 3 times larger than a normal unit-triangle (see section 7). Therefore for series (c), equation (12) becomes:

$$D = 3n_H * \sqrt{(30C/\pi)} \qquad \text{(equation 13)}$$

The value of C is given by Hauser et al. [16]; under conditions where membranes are liquid-crystals $C=.25nm^2$. By substituting this value, the following equations are obtained for the effective diameters, D, of vesicular monolayers (all diameters are given in nm):

for series (b): $D = 3n_H * \sqrt{(2.5/\pi)}$ (equation 14)

for series (c): $D = 3n_H * \sqrt{(7.5/\pi)}$ (equation 15)

For real vesicles, D is the diameter of their inner monolayer. The diameter, D(V), of the outer layer (as discussed in section 7 and demonstrated in Fig. 11) is given by:

$D(V) = D+2(5+1) = D+12$ (equation 16)

If in this equation D is substituted by (14) and (15) the following equations result for the theoretical diameters of vesicles of biomembranes:

for series (b): $D(V) = 3n_H * \sqrt{(2.5/\pi)} + 12$ (equation 17)

for series (c): $D(V) = 3n_H * \sqrt{(7.5/\pi)} + 12$ (equation 18)

In Table 2 the results obtained with these equations are presented

In Table 3, theoretical diameters are compared with the values obtained by electron microscopy of vesicles for neurotransmitters and hormones.

To demonstrate for in vivo vesicles that the icosahedron-model is correct, a statistical analysis of the data in Table 3 is required. The results of this analysis are given in Fig. 16, and is based on the hypothesis that the diameters of vesicles are random or that at least no simple relations exist between the diameters. In other words their continuous distribution of sizes is arbitrarily divided by the theoretical values of Table 2.

The curve shows that the maximal frequency occurs at an average deviation of 4.3% (at this maximum the frequency is about 45000 times). Furthermore the number of averages with a value of 1.8% (the value determined from Table 2 for in vivo vesicles) or smaller is only about 200 per million (p=0.0002).

Since the hypothesis of randomness cannot be reconciled with p=0.0002 it must be concluded that the result for the diameters of in vivo vesicles in Table 3 is very significant and that the sizes of in vivo vesicles are not random but (as explained by the theory) determined by two geometric series ($2^n$ and $3*2^n$), corresponding with the two series for the size of hexagons.

The conclusions in this article do not depend on papers with only one single diameter but on 3 sets with at least three. This makes it more difficult to manipulate the data by leaving out one or more diameters that do not fit the theory. Despite these precautions still a justification of the choice of the in vivo vesicles in Table 3 is required.

Although with the 19 diameters of Table 1 the statistical analysis would even be more convincing, only 12 of the 19 diameters in Table 1 were used in Table 3, and 7 values were deleted. In the first place the 6 diameters for hormone vesicles from rat [11] were deleted for the following reason. The authors determined two diameters for every vesicle: one for the vesicles observed in situ and one for the same vesicles after they were isolated. In this way two sets of diameters were obtained: one set are diameters measured in situ and the other set the diameters of the isolated vesicles. In Table 1 these two sets of diameters were combined to one single set by taking for every type of vesicle its average diameter. Since there is an uncertainty in the actual diameters they were not used in the statistical analysis. The diameter of human LTH-vesicles

[14] was also deleted. Without statistical analysis it can be concluded that these vesicles do not fit into the model because their diameter of 559.9 nm deviates too much (6.1%) from its theoretical value. This diameter could be the evidence required to proof that the icosahedron-model is not correct! A deviation of about 6% cannot be tolerated, unless it can be explained. The relatively large diameter of human vesicles for LTH must be attributed to their elliptical shape, as reported by the author [14]. It is well documented as was discussed previously [4] that artifacts can be introduced by our manipulations. One of these artifacts is that the real dimensions of particles can become distorted. For instance spherical vesicles can become ellipsoids. The relevant quantity is the surface area of a vesicle and when its shape is spherical we need only one single linear dimension, the radius, for exact calculations. However, if a vesicle deviates from the spherical shape (like the one for human LTH) we need more than one linear dimension to calculate its surface area.

12. Discussion.

The solution for the problem as given in this paper seems to be correct. In the first place it was demonstrated that the diameters of in vivo vesicles could be calculated with high accuracy (see Table 3 and Fig. 16). In literature one looks in vain for another equation to calculate all diameters that occur in vivo. In the following it is shown that also other conclusions, derived from the model, are realized by in vivo vesicles.

What is the driving force behind the process that leads to discreteness of the size of vesicles? Because ultimately all processes also in living nature, are governed by thermodynamics, it is assumed that the solution of the problem must be sought in the statistical mechanics of phospholipids. Phospholipids are an essential component of membranes, have a low molecular weight, and form the majority of molecules. The total number of configurations of a vesicle with a certain $n_H$-value is the sum of all configurations that contribute to the partition function. Since also the formation of bilayers is induced by phospholipids, the largest number of configurations of vesicles is given by the formation of regular hexagons on both monolayers. It must be kept in mind that the total number of configurations of the phospholipids of a vesicle is not simply the sum of the configurations of its two constituent monolayers and that for the calculation of the partition function of the total bilayer, the energy of every configuration must be calculated. The latter energy in turn, is determined by factors like the translocations of molecules, the rotations of arrays of molecules, the distribution of vacancies and the interactions of the headgroups. The author is not able to solve the quantitative aspects but knows that these calculations must be performed to proof conclusively that the model is correct.

Additional experimental proof for the correctness of the model is given by the following analysis of the data. First the difference between the surface areas of the two monolayers of a vesicle as a function of its diameter was calculated. Starting with diameters, D, of the inner monolayers, and taking a constant distance, $\Delta$, between the two monolayers, using the equations $\pi D^2$ and $\pi (D+\Delta)^2$, the surface areas of respectively the inner and outer layer can be calculated. The value $100*((D+\Delta)^2 - D^2)/ D^2$ (%) or $100*((1+\Delta/D)^2 - 1)$ (%) is defined as the relative difference between the surface areas of the two monolayers. Since $\Delta$ is always much smaller than D, the relative difference is approximated by 100*2Δ /D (%). Irrespective of the model used, i.e. irrespective of the value of Δ , this relation holds. It is evident that the larger the diameters, D, the smaller the relative difference. In the following it is assumed that the icosahedron model is correct and therefore Δ = 4 nm; and the relative difference equals 800/D(%).

The diameters, D, of the inner layers of in vivo vesicles are determined by subtracting 12 nm from their given outer diameters. Of the four smallest in vivo vesicles (see Table 3) the diameters of the inner layers are 32.3, 38, 56.8, and 66.7 nm. A fifth diameter of 54.8 nm (see Table 2) not existing in vivo is a theoretical value; the diameter of the inner layer of this theoretical vesicle should be 42.8 nm.

With the relative difference = 800/D (%) the following percentages between the two monolayers of these 5 vesicles can be calculated: 25, 21, 19, 14 and 12 %. One of the most important implications of the icosahedron model is, that the two layers of a bilayer are supposed to have the same number of molecules of diglycerides. It is not allowed to ignore the hard fact that a large relative difference of respectively 25, 21 or 19% can hardly be reconciled with the same number of molecules of diglycerides in both layers. On the other hand it is also not allowed to ignore the equally hard fact that the second term of series (b), a vesicle of 54.8 nm, with a corresponding inner layer of 42.8 nm, is only a theoretical vesicle. For the second diameter of series (b), D(2), it was mentioned already 25 years ago, that "..it is striking that a vesicle corresponding to D(2) has never been observed." [4]. In other words vesicles with a diameter of about 54.8 nm could not be found, neither in vivo nor in vitro, and a solution for this finding could not be given. In the following it is shown why a vesicle of 54.8 nm cannot exist.

All experimental data can be explained if it is assumed, that the size of all vesicles is described by the icosahedron model. A relatively large difference between the surface areas of the two monolayers, seems to be the only prohibiting factor for the formation of bilayers. Therefore vesicles of 54.8 nm cannot exist because a difference of 19% between the two monolayers is apparently too large.

If formation of vesicles of 54.8 nm is not allowed because the difference between their monolayers is 19%, why two smaller vesicles with still larger differences of respectively 21 and 25%, do exist?

An answer to this question can be found in the values of the three theoretical diameters of respectively 44.1, 49.1 and 54.8 nm given in Table 2. The first two diameters correspond with those of the two smallest vesicles observed in vivo; the third diameter is the theoretical value not found in vivo. Note that the differences between these vesicles, of respectively 5 nm and 5.7 nm, are also the differences between their inner monolayers. It must be concluded that the inner monolayer of 54.8 nm instead of functioning as an inner monolayer can also function as the outer monolayer of a vesicle of 49.1 nm because there is only a difference of 5.7 nm (= 42.8-37.1 nm) instead of 4 nm between its two monolayers. The same holds for the inner monolayer of the vesicles of 49.1 nm; it is only 5 nm larger than the inner monolayer of 44.1 nm. Therefore the inner monolayer of 49.1 nm has still another function: it is also the outer monolayer of 44.1 nm.

The first term of series (b) and the first term of series (c) are stable, despite the large relative difference in surface area between their monolayers; it are hybrids because their inner monolayers and outer monolayers have a different lattice. Apparently also the diameters of the smallest vesicles can be calculated with the icosahedron model despite a difference in the number of molecules of fatty acyl chains of their monolayers.

The following range of diameters is proposed: A) very small vesicles, B) very large vesicles and C) intermediate vesicles. These 3 groups of vesicles are now discussed.

A) Very small vesicles. As discussed above the two smallest vesicles would not have existed, like the vesicles with 54.8 nm, if not for these vesicles, outer monolayers with a different lattice but with about the required diameter, can be constructed.

B) Very large vesicles. A small difference in cross-sections between the chains of the two monolayers of very large vesicles could explain a small relative difference in surface area. For instance an asymmetry could exist in the fatty acyl chains of respectively the outer and inner monolayers.

C) Intermediate vesicles. In these vesicles the difference in surface area between monolayers could be too large to be explained by a difference in cross-section only. However, it cannot be excluded that in these vesicles the monolayers also differ in the size of their hexagons. The inner monolayers could have 20 of the largest hexagons while the outer monolayers could, with the same number of molecules of diglyceride, have smaller hexagons and therefore more vacancies and therefore a larger surface area.

Of the 16 theoretical diameters, that are given in the interval between 44.1 and 457 nm (see Table 2), 11 diameters of in vivo vesicles are represented in Table 3. Of the 8 theoretical diameters calculated for series (b), 6 diameters of in vivo vesicles are represented in Table 3. Vesicles with about 54.8 nm cannot exist, leaving vesicles with about 268.9 nm as the only ones of series (b) that could have been realized but are missing. Of the 8 diameters in series (c) that could have been realized, three diameters with values of about 86.1 nm, 160.3 nm and 308.7 nm are missing in Table 3.

In addition to analyzing the difference in diameter between the two monolayers of the same vesicle, as was done above, also the difference between the diameters of two inner monolayers adjacent in size can be calculated. These theoretical intervals must then be compared with the measured ones.

Provided the icosahedron model is correct, the diameters of the inner layers fit in two series. One series is dictated by the numbers 2, 3, 4, 6, 8………(see series the (a)), while the terms of the other series are $\sqrt{3}$ larger and are represented by the numbers $2\sqrt{3}, 3\sqrt{3}, 4\sqrt{3}, 6\sqrt{3}\,8\sqrt{3}$……When the terms of these two series are combined and the diameters arranged according to size, the diameters are represented by: …….4, $3\sqrt{3}$, 6, $4\sqrt{3}$, 8, $6\sqrt{3}$……For the part of the combined series shown here, the theoretical interval between the first two terms is: $100*(3\sqrt{3} - 4)/4 = 30\%$ while the following three intervals are 15%. If the icosahedron model is correct, the intervals between the inner layers form the repeating sequence 30, 15, 15 and 15%.

Of the following 7 pairs of in vivo vesicles given in Table 3 it is possible to determine the intervals between the diameters of two adjacent inner monolayers. Between the diameters 44.3 and 50.0 nm the interval is 17.6%, between 68.8 and 78.7 nm it is 17.4 %, between 95 and 125 nm it is 36.1 %, between 125 and 135.5 nm it is 9.3%, between 181.6 and 226 nm it is 26.2% and between 356.8 and 452.9 nm (twice!) it is 27.9%. The interval between two measured diameters must always deviate from the theoretical value. If however after careful measurement of a certain diameter, the deviation from its theoretical value is about 22% (the average between 15 and 30%) it must be concluded that the model is not correct! All arguments in favour of the model are then based on coincidence and show clearly the speculative character of the model. If however the model is correct we must require that the average of all intervals below 22% must be about 15% and above 22% about 30%. The averages of the two groups of intervals of in vivo vesicles given above are 15% and 30%. Also this results is an argument that the model is correct.

If the icosahedron-model is correct, and vesicles of about 54.8 nm are not detected in vivo, between the two inner monolayers of vesicles with 50 nm and 68.8 nm a large interval of 50% must exist, while all other intervals are about 15% or 30%. If a vesicle with about 54.8 nm is detected the model cannot be correct. If on the other hand the four diameters missing in Table 3, or even one of them, is given in literature, it would be another indication that the model is correct.

It must be realized that even if hexagons cannot be observed, it does not exclude the preferential formation of hexagons on vesicles. The statement "preferential formation of hexagons" means that the formation of hexagons with a certain size contributes to the partition function of phosphodiglycerides but does not exclude the almost simultaneous formation of hexagons with another size. Therefore in general, due to the large mobility of the phospholipids, one could look in vain for the occurrence of hexagons on membranes.

If for some reason (for instance due to the incorporation of proteins) in part of the cell, the mobility of phospholipids is inhibited and the formation of hexagons restricted, a configuration could prevail that does not necessarily contain hexagons. This probably explains the continuous distribution of the size of cells in the S-phase during cell division, although the formation of even-honeycombs and therefore a large mobility on only a small part of the membrane, remains possible. Incorporation of the cytosceleton into "strategic" points of the outer membrane of a cell, could inhibit free movement of the phospholipids, prevent the formation of a large number of different hexagons. and so enable the continuous distribution of the size of cells in the S-phase.

Vacancies induce a higher degree of freedom for molecular rotations [23]. In liquid crystals, like lipid membranes, rotations are actually cooperative jumps with a very high frequency ($10^9$ per sec; cf. [15]). It can be imagined that one single vacancy in front of an array promotes the cooperative rotation of all molecules in this array while after rotation the vacancy appears at the other end of the array. In this way vacancies diffuse very rapidly over large distances.

An important argument against the model is, that it is difficult to accept, that the size of biological vesicles with considerable compositional heterogeneity is predicted on the basis of the cross-section of a single fatty acyl chain. But instead of looking at the compositional heterogeneity, it is more rational to consider a feature that all acyl chains in a monolayer have in common. The part of a fatty acyl chain formed by its first 7 C-atoms, is identical for all chains and it is this part that creates the layer of liquid crystal where the model must be applied.

The phospholipids in even honeycombs have a very high mobility, induced by vacancies, and are continuously mixed. Therefore compositional heterogeneity is not an argument against the model; due to the heterogeneity, the number of permutations and therefore the number of different configurations is increased enormously. In other words heterogeneity is an advantage and it could even be essential to take the permutations into account. Without vacancies i.e.without mobility, like in odd honeycombs, there is only one configuration.

As explained, vacancies diffuse rapidly by rotations with high frequency, and therefore two vacancies will certainly meet and form a pair that disappears when replaced by a molecule of diglycerides. On the other hand it is certain that without vacancies there is no mobility and without mobility no permutations. Are even-honeycombs the only configurations with vacancies that stay single and that cannot disappear by forming pairs? Let us suppose that without permutations the number of different configurations of a vesicle is so small that there would not be discreteness. Then it can be understood that only with even honeycombs, via permutations all possible configurations are realized, resulting in discreteness of the size of vesicles.

Another aspect is the determination of the surface areas of very large vesicles. Because it was shown that not only small in vitro vesicles but also very large vesicles of cellular dimensions fit into two geometric series [34], those who are mainly interested in biology (like the author), are willing to accept that the discreteness of in vivo vesicles includes cells. Cells fit in the two geometric series only at mitosis and not in other phases of the cell cycle because cells in mitosis have no cytoskeleton [5]. If the measured diameters of mitotic cells indeed correspond with the calculated values of very large vesicles, one has to accept that the model can be applied to the whole range of vesicles. Then it is legitimate to conclude that in vivo over the whole range of sizes, vesicles are discrete.

The author hopes that the experimental evidence for the existence of discreteness is accepted by those who are able to perform the statistical mechanics of the problem. If the theory can be proved unequivocally to be correct, its ramifications for cell biology, can and will be investigated more thoroughly. It was already shown how discreteness of cells can explain the formation of blood cells from stem cells in bone marrow [5].

Especially the discreteness of a group of spherical particles like nuclei must be investigated. If it can be calculated that also the size of nuclei (the organelles with the genetic material!) is discrete, this fact would have important consequences for biology [5].

Another problem (perhaps the most important one!), that could be solved after fundamental calculations on the statistical mechanics of phospholipids are performed, concerns investigations of nonspherical membranes. For instance it must be realized that a great part of cells of brain have plasmamembranes that are not spherical. Is it possible, that knowledge obtained from spherical membranes can be applied to nonspherical ones?

Diameters of vesicles can be determined by three completely different, independent methods: two experimental and one theoretical. Diameters can either be determined directly by electron microscopy or derived from the mass determined by analytical centrifugation. They can also be derived from the theoretically determined surface areas. The remarkable agreement between the three methods (they give almost the same diameters) makes it likely that each of them is valid.

The model for membranes is based on the properties of phospholipids but it is required to investigate whether the model proposed here is not prohibited by other components like proteins.

One of the essential functions of proteins in the stabilisation of membranes can be understood as follows. In monolayers with a normal lattice every vacancy with high energy is surrounded by 6 lattice-points also with high energy, because they are derived from the 6 chains of 3 molecules of phospholipids that make angles of 60°. In bilayers like biomembranes both monolayers contain vacancies because both are covered with honeycombs and the energy of biomembranes is increased accordingly. It is assumed that in a biomembrane a channel can be formed by a vacancy in one monolayer juxtaposed with a vacancy in the other monolayer. The 6 molecules of phospholipids that form the channel, can be replaced by chains of protein. By the presence of proteins in vesicles two vacancies (one per monolayer) with high energy are transformed into one single channel supposedly with lower energy. Channels that are formed by proteins can be used for communication between cells, as illustrated in the following example.

In certain tissues junctions can be formed where two cells, i.e. two bilayers, meet. For our discussion so-called gap junctions are important. While most of the material isolated after fragmentation of membranes forms vesicles, the material that contains the gap junctions is flat and does not contain spherical bodies [10]. Isolated gap junctions are two-dimensional flat fragments, built up of units called connexons. Where two cells meet, connexons of the gap junctions span the pair of bilayers, including the gap between the two cells; they provide the connection for intercellular communication. The following microscopic characteristics were observed: every connexon has a six-fold axis of rotation showing that it has a hexagonal structure [1] and that in turn all connexons form together a triangular lattice [10]. The formation of gap junctions and the properties of connexons can be explained if it is accepted that they result in the first place from the formation of honeycombs in the matrix formed by the phosopholipids (cf Fig.14), and by the concomitant replacement of diglycerides with high energy around vacancies, by proteins.

Membrane proteins form two classes. One type, the transmembrane proteins span the whole bilayer while the others are integrated into one of the monolayers. Certain lipoproteins in monolayers, could for instance function as filler molecules by occupying the vacant centers in a centered lattice.

Arguments why proteins are preferentially located in vacancies are given by Jain [21] who calls them defect sites, and "binding sites" for proteins.

This paper does not pretend to give an exhaustive treatment of an extremely complex problem. It was written in the first place to draw the attention to the discreteness of in vivo vesicles. Therefore, in addition to the evidence already given, also other facts that demonstrate the existence of discreteness have to be mentioned. First of all Kelly [22] working with synaptic vesicles, wrote: "We now know from electron micrographs that synaptic vesicles are indeed homogeneous in size, perhaps the most homogeneously sized vesicles in biology" and furthermore: "What was truly amazing was the narrow size distribution of the packets". The conclusions of the author are not only based on the determination of diameters but in the first place on very accurate determination of the neurotransmitter signal by electrophysiological measurements. If this author had studied the size of other vesicles as carefully as those of neurotransmitters he would certainly have included all vesicles in his statement. Like vesicles for neurotransmitters also the vesicles for hormones are homogeneous in size. The fact that vesicles for neurotransmitters belong to the small ones cannot be used as an argument for a special position. Even the smallest of these vesicles have still the relatively large number of about 15000 molecules. In addition their surface areas correspond formally with those of icosahedrons who are terms of series (b) and series (c). Finally, the statistical analysis of Table 3 excludes randomness of size of vesicles both for neurotransmitters and for hormones.

If the laws of thermodynamics are violated by the requirement of the preferential formation of hexagons by phospholipids, the icosahedron-model of vesicles cannot be correct. Therefore calculations of the energy and entropy of bilayers must be performed. For investigations of the free energy of membranes it is perhaps possible to compare different configurations of phospholipids of membranes with the same size; configurations with hexagons must have an advantage. It must then be kept in mind that the smallest vesicles that are formed in vivo have $n_T=36$ and that vesicles with $n_T=15$, like those in Figs 14 and 15, could be too small for reliable calculations.

Acknowledgements

In the first place I`m obliged to Prof. D. Frenkel not only for his help to publish this paper but also for his understanding and encouragement. Further I have to thank: Prof. H.V. Westerhoff for his advice and critical reading of many versions of the manuscript; Dr R. van Dantzig for fruitful discussions and drawing my attention (already 20 years ago!) to equation (7) and to a regular polygon as a possible model for vesicles; Prof. C.G. Figdor for critical reading of the manuscript; Dr J. Boom for stimulating discussions and construction of figures; Dr K.S. Kits for providing me with the information of synaptic vesicles; Dr P. van Kasteren for performing the statistical analysis; Dr E. Wattel for the construction of Fig. 12, and to the library of The Netherlands Cancer Institute for providing the literature.

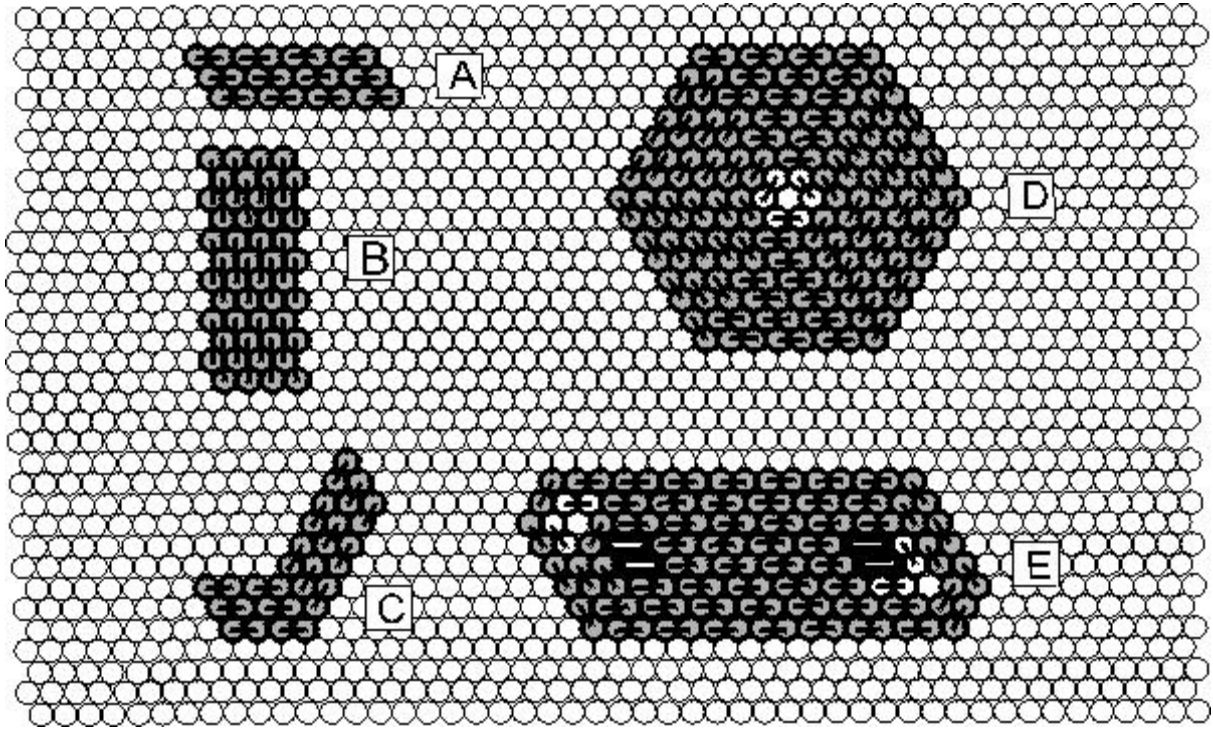

**Fig. 1**

*The fatty acyl chains (circles) are hexagonally close-packed. The predominant configurations are closed linear arrays that make accidentally angles of 120°.*

*A) Diglycerides are represented by 2 black circles connected by a black bar. This configuration is called the normal one. The hexagonal close-packing permits three directions for the arrays; in this configuration only the horizontal one is used.*

*B) The molecules form linear arrays in the vertical direction This configuration is called the "perpendicular" one because the orientation of the molecules is perpendicular to the normal configuration.*

*C) The linear arrays form bands that make angles of 120°; in this configuration two of the three directions are used.*

*D) The arrays form regular hexagonal structures with in the middle a vacancy, a point without a chain. A vacancy is formed by three molecules (with open circles connected by a black bar) that make angles of 60°. In this configuration all three directions are used.*

*E) The arrays form irregular hexagonal structures. This leads to angles of 60° (molecules with open circles connected by black bar) and/or to more than one vacancy and/or to arrays that are not closed but have open ends (molecules represented by two black circles connected by white bars).*

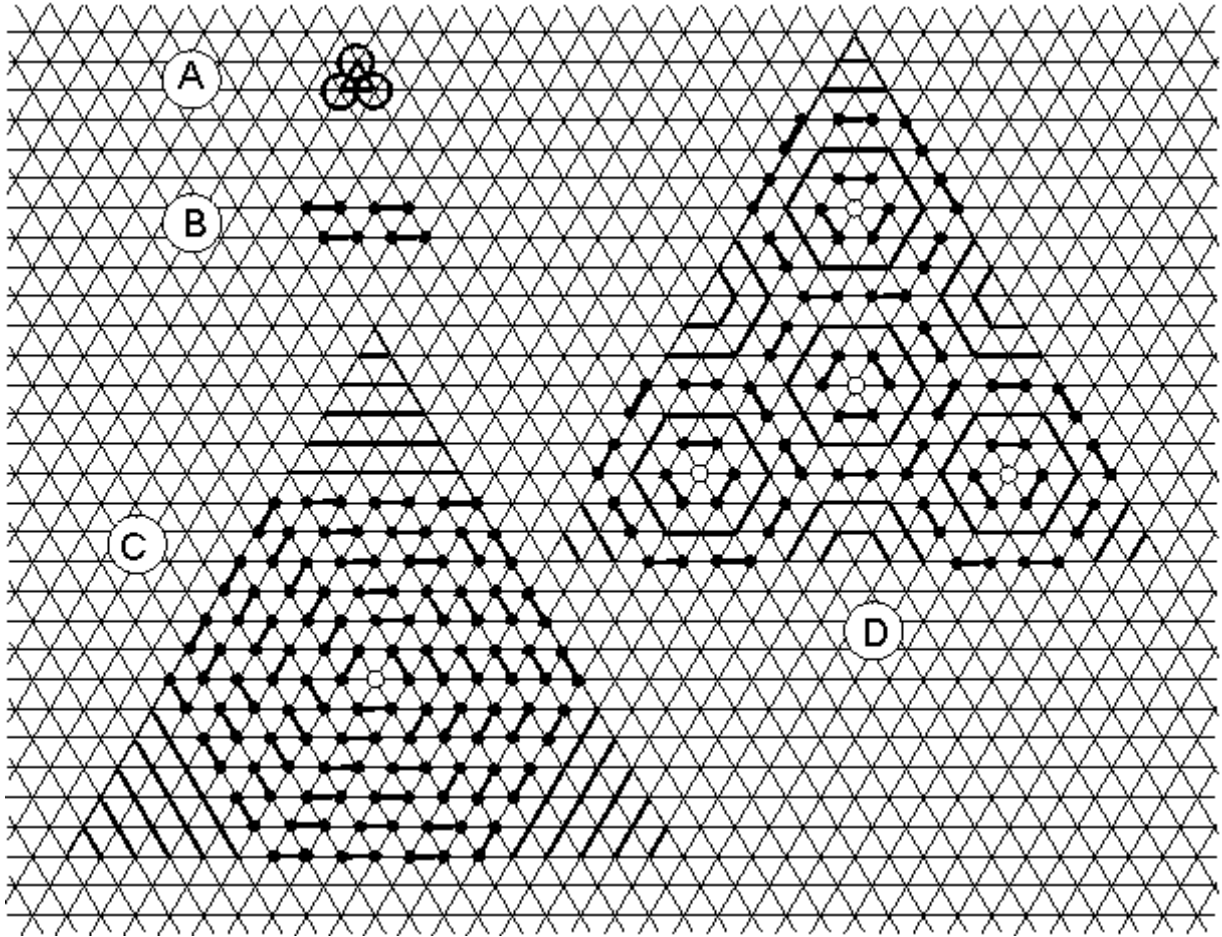

**Fig. 2**

*The hexagonal close-packing of the chains of Fig. 1 is represented by a triangular crystal lattice.*

*A) A unit-triangle of the triangular crystal lattice. The chains are situated in the corners of the unit-triangles. Every chain of the lattice is divided by six unit-triangles. The lattice-constant, d, is given by a side of a unit-triangle.*

*B) A molecule of phospholipids is represented in the lattice by two black dots connected by a bar.*

*C) The size of a triangular monolayer is defined by $n_T$, the number of lattice-constants of one of its sides; in the figure $n_T=18$. The size of the largest regular hexagon, H, that can be constructed in T is defined by $n_H$, the number of lattice-constants of one of its sides; in this figure $n_H=6$. A hexagonal structure is formed by nesting of the hexagon; this structure is completely filled except for the central vacancy. The part of T not covered by the hexagon is covered by linear arrays represented by black lines.*

*D) Part of T is covered by a honeycomb with $n_{hon}=3$; the rest is covered by arrays represented by black lines. To accentuate the honeycomb, one of the nested hexagonal arrays in every hexagonal structure is represented by a hexagon and not by molecules. In this example with $n_{hon}=3$ the number of hexagons of the honeycomb that have one side coinciding with one side of T, amounts to f(n)=2 (see equation (4)).*

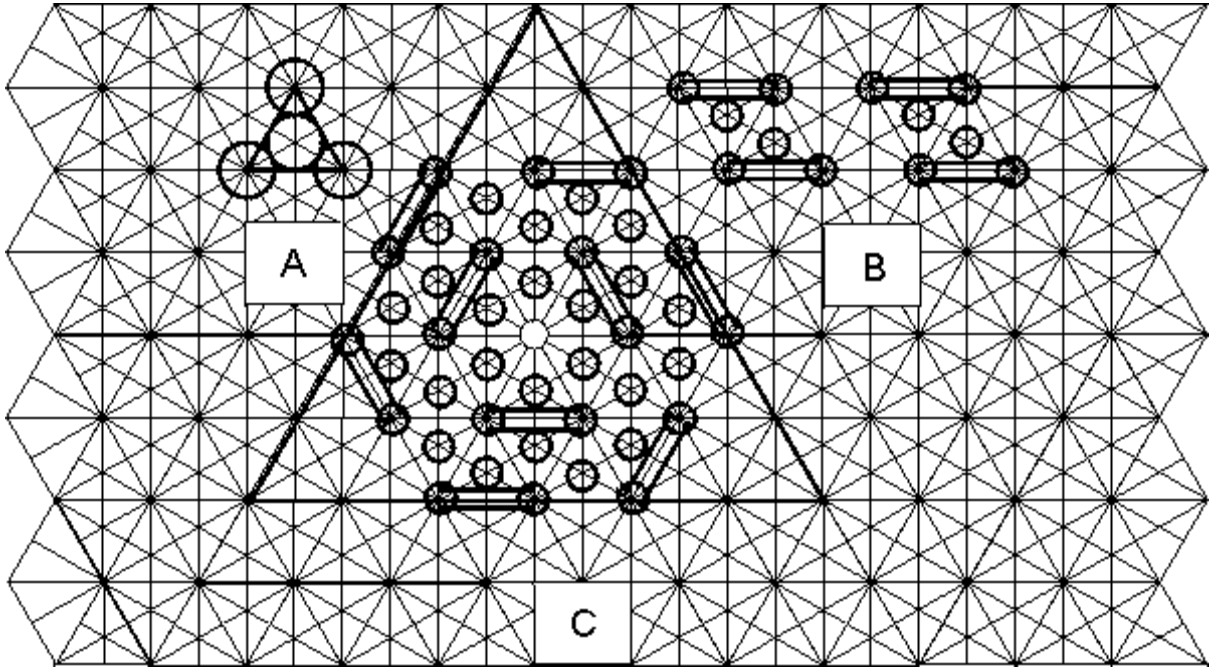

**Fig. 3**

*A centered triangular lattice is constructed from a normal one. The directions of the lines in a centered lattice are perpendicular to the three directions in a normal lattice (see Fig. 1). The corners of centered unit-triangles become visible as black dots because there six lines cross each other.*

*A) A centered unit-triangle. In a centered lattice a chain in a corner (derived from a diglyceride) is shared by six centered unit-triangles, while a chain in a center (derived from a molecule with only one single chain) is situated completely within a unit-triangle.*

*B) The lattice points in a centered lattice are occupied either by diglycerides (two circles connected by two lines) or by single chained molecules (single circles). A centered lattice is constructed by replacing two adjacent linear arrays of diglycerides in the hypothetical "perpendicular" lattice of Fig. 1B by single-chained molecules. The latter form the centers.*

*C) Centered unit-triangles can be treated like normal unit-triangles; by using the three possible directions they form hexagons (cf. Fig. 2C).*

*Vacancies in the middle of hexagons (here represented by an open circle) are surrounded by six centers. Note the difference between vacancies in a normal and in a centered lattice.*

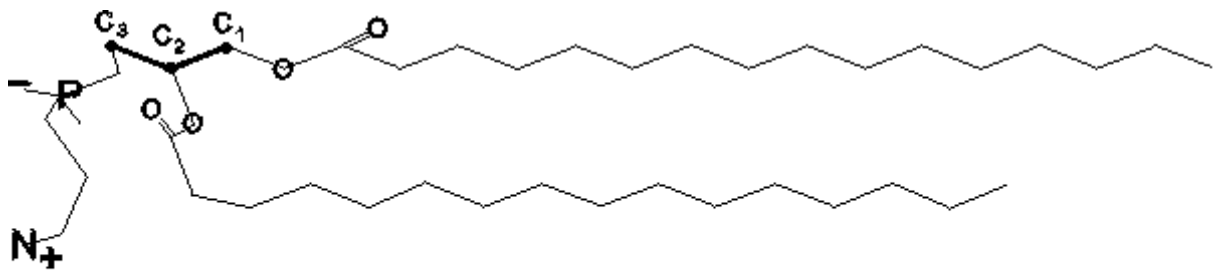

Fig. 4
Schematic representation of the molecular structure of PL. The glycerol moiety, indicated by its three C-atoms, is in one line with chain 1. Chain 2 bends of at its second carbon atom, to become parallel to chain 1. The dipolar headgroup is approximately parallel to the surface (cf [36]) while the chains are perpendicular to the surface.

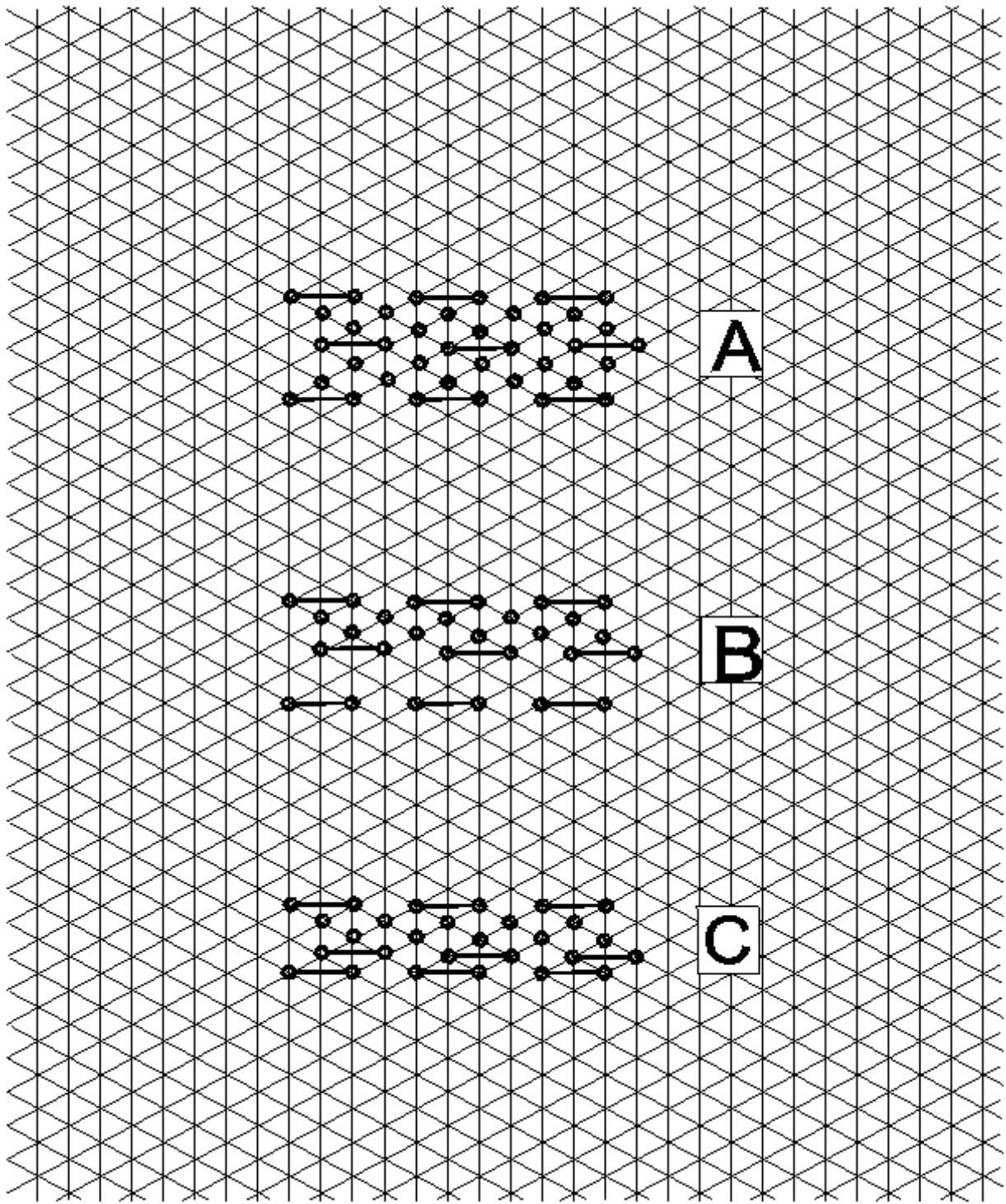

**Fig. 5**

*This figure deals with a centered lattice (see Fig. 3B). although only the normal lattice is shown. The corners are occupied by chains from molecules of DPPC (circles connected by lines) while the chains in the centers represent PA (single circles).*

*A) All centers are occupied by PA.*

*B) Two arrays for centers between two arrays of PL-molecules are alternately occupied by PA or empty.*

*C) The situation of B after condensation.*

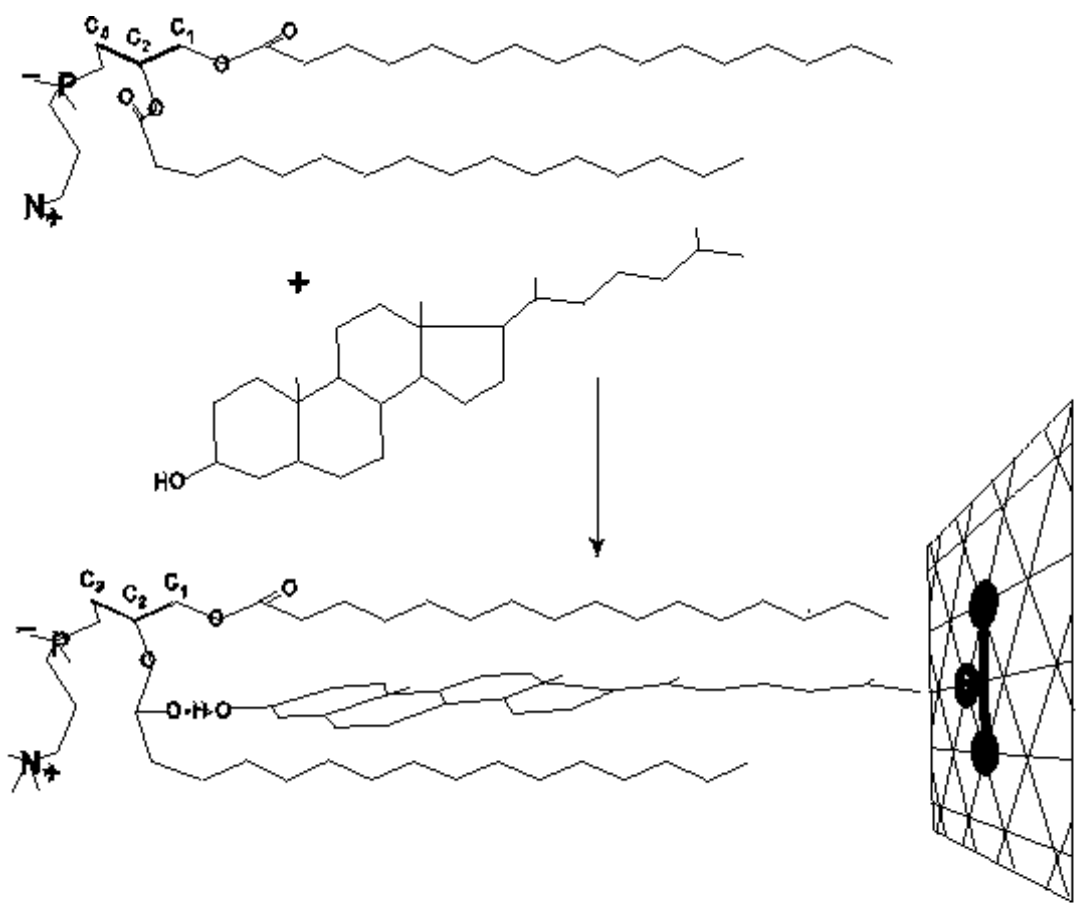

Fig. 6
Formation of a 1:1 complex of Chol with PC takes place by intercalation of the sterol group between the two chains and a reaction of the OH-group with the CO-group of chain 2. The two chains occupy corners while the hydrocarbon chain of the Chol is placed in a center.

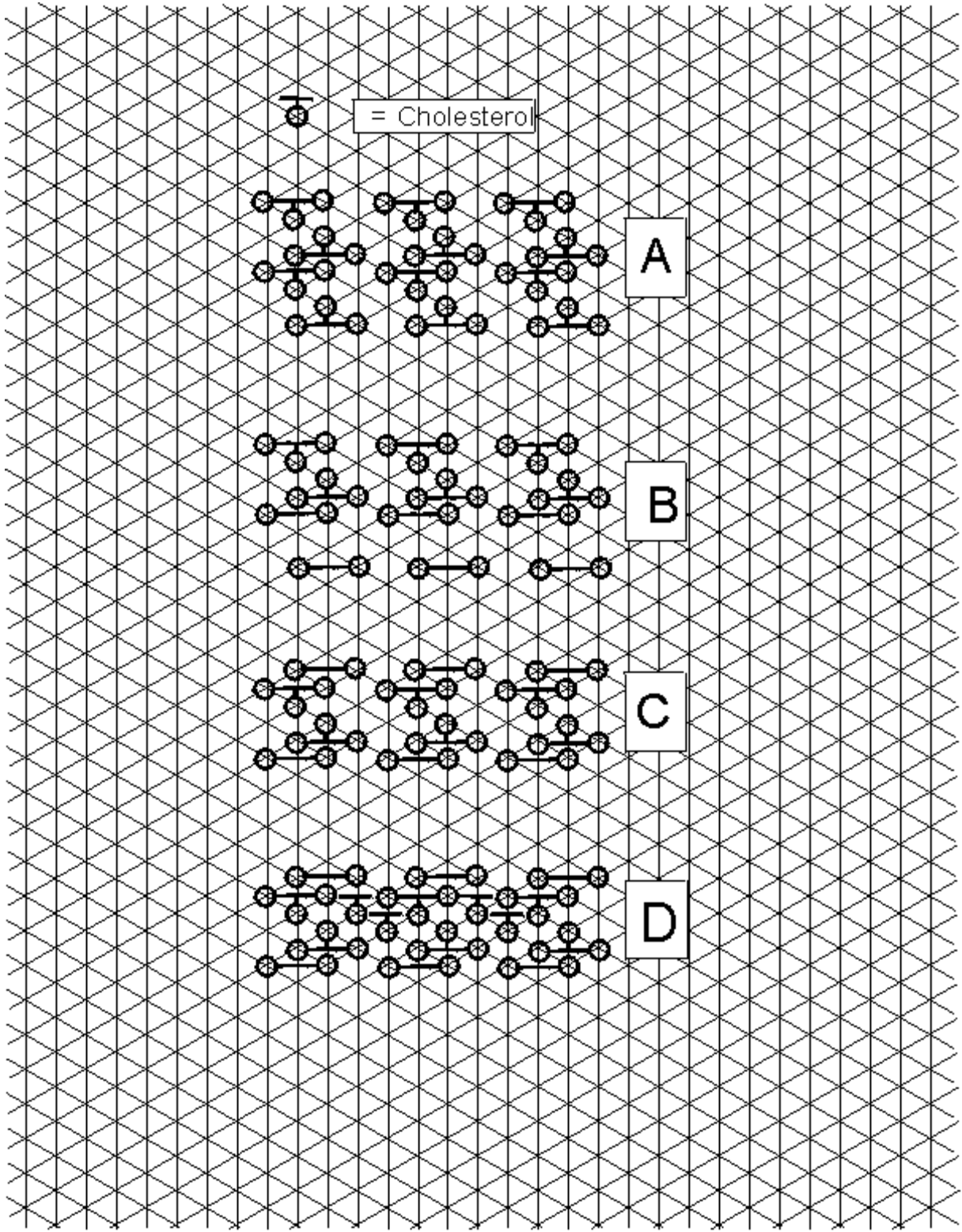

**Fig. 7**

*The formation of a centered lattice with PC and Chol.*

*A) The lattice formed by linear arrays of 1:1 complex (cf. Fig. 6) shows the following characteristics. The orientation of the linear arrays of diglycerides indicates that the centered*

*lattice is derived from the perpendicular lattice of Fig. 1B; the chains of cholesterol occupy the centers. The lattice has still pairwise vacant points.*

*B) Two linear arrays of 1:1 complex alternate with two arrays of pure PC.*

*C) The situation of B has too many vacancies; condensation must follow.*

*D) A further reduction of vacancies. In addition to functioning as an intramolecular spacer in a 1:1 complex (cf Fig. 6) in the lattice Chol could perhaps also be present as an intermolecular spacer between two PC molecules and occupy the empty lattice points in A).*

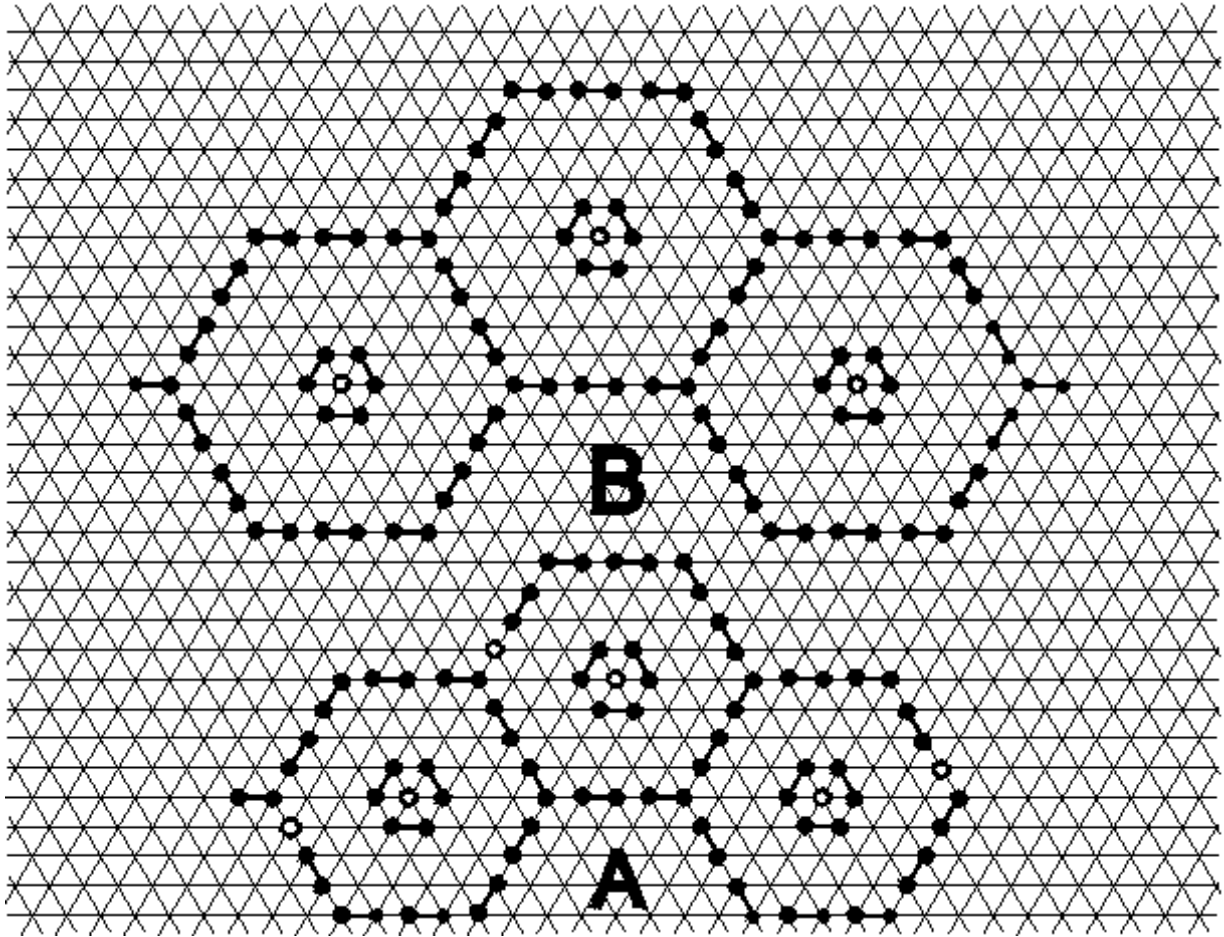

**Fig. 8**

*The difference between honeycombs with respectively even or odd $n_{hon}$-values, is illustrated with 3 hexagons of each group.*

*A: Honeycombs with even values (here $n_{hon}=4$) with one vacancy per hexagon are designated as even-honeycombs.*

*B: Honeycombs with odd values (here $n_{hon}=5$) are designated as odd-honeycombs and have no vacancies.*

*The nested, hexagonal arrays were omitted, except around vacancies.*

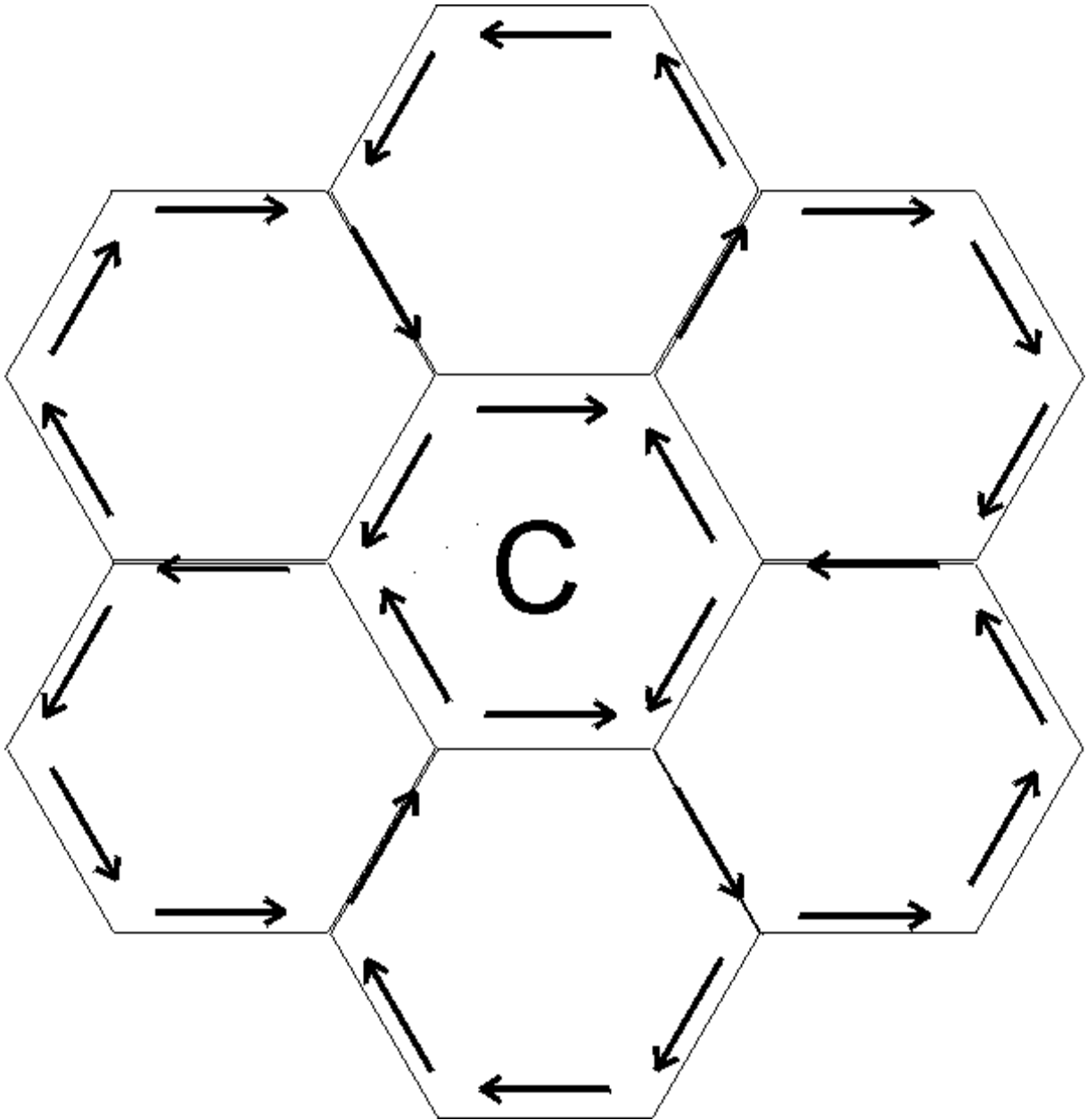

**Fig. 9**

*This figure shows the direction of dipoles in an odd-honeycomb.*

*In six hexagons the direction of the dipoles is either clockwise or counter-clockwise. This leads to a disruption of the cooperative interaction of the dipoles in the central hexagon, resulting in an instable honeycomb.*

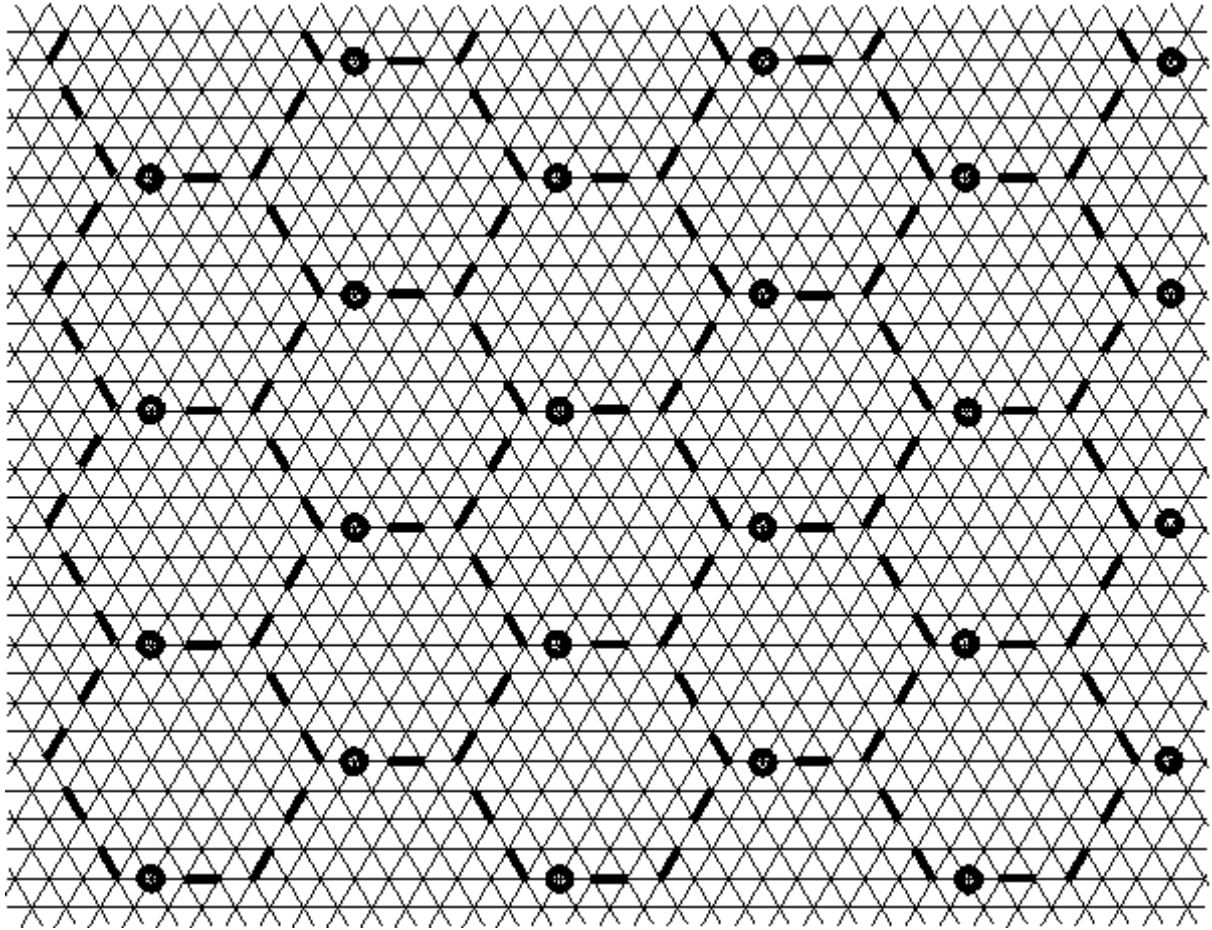

**Fig. 10**

*The representation of diglycerides is reduced to a bar connecting two lattice points.*

*In an even-honeycomb the number of locations for the vacancies (open circles) is relatively large. In the configuration shown here (one of the many that are possible) every hexagon has one vacancy since it has two vacancies that must be divided with its neighbours.*

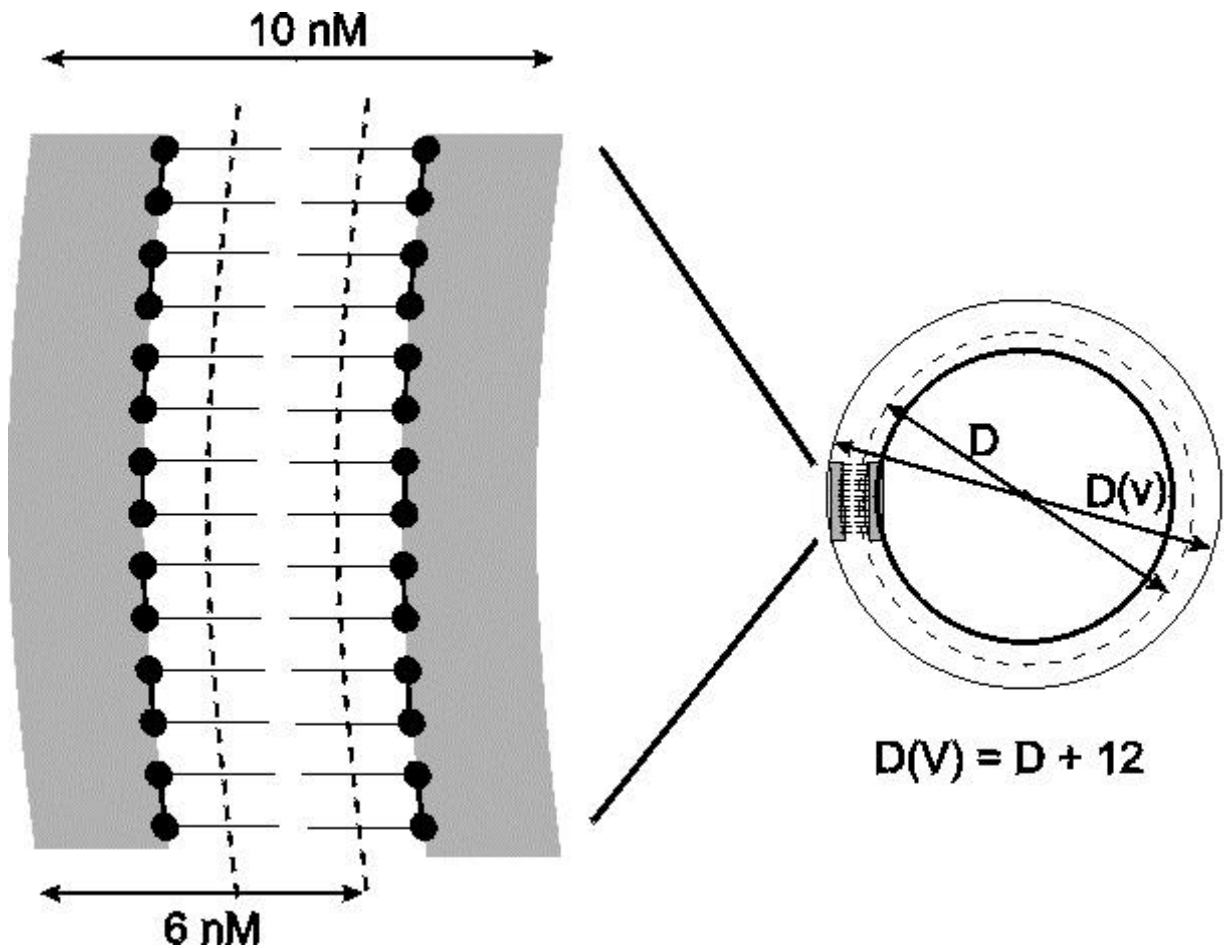

**Fig. 11**

*A cross-section of a biomembrane.*

*A biomembrane (10nm) is formed by a PL bilayer of about 5nm, sandwiched between two protein layers (thick gray layers) each of about 2.5nm.*

*The planes for the following diameters of vesicles are indicated:*

*D(V)=outer diameter of vesicle; D=effective diameter of the inner monolayer represented by a broken line; also the effective diameter of the outer monolayer is represented by a broken line. The distance between the two broken lines is 2nm and therefore the difference between the effective diameters of respectively the inner- and the outer layer is 4 nm.*

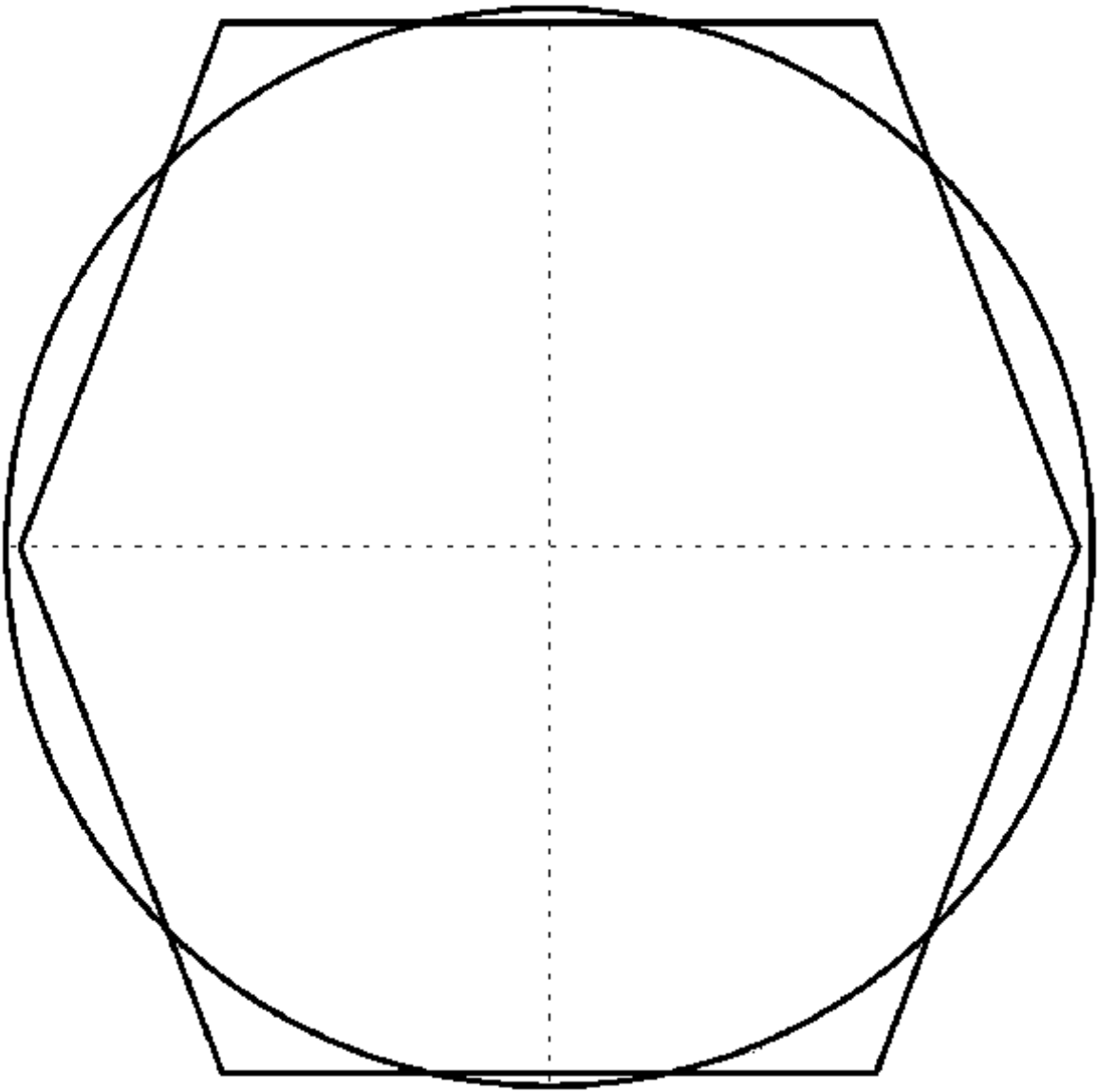

**Fig. 12**

*Cross section through the center of an icosahedron and the sphere with the same surface area.*

*This cross section through 4 of the twelf corners of the icosahedron shows the largest deviations that can be expected between an icosahedron and its sphere. It is well known that an icosahedron with truncated corners is almost a perfect sphere and therefore the deviations in all other cross sections are even much smaller.*

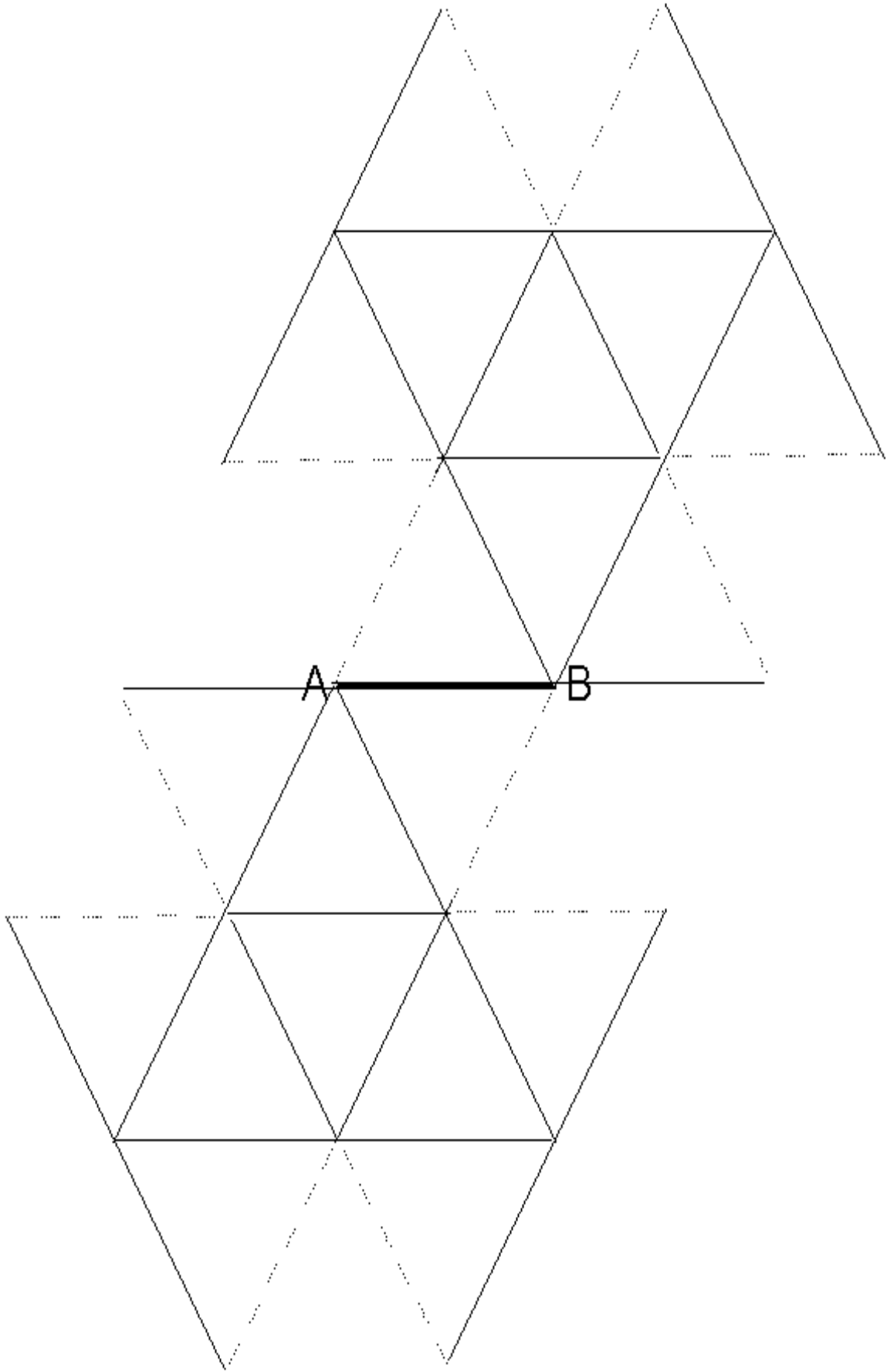

**Fig. 13**

*Icosahedrons are represented in a plane. The 20 triangles of an icosahedron are divided in two identical parts. The two parts are linked via a common side (A-B) between two triangles. A three-dimensional icosahedron can be constructed from this two-dimensional representation, by joining pairs of triangles with a broken side and folding the two parts along the line A-B.*

*This procedure for the formation of an icosahedron can be checked with (a magnification of) this figure.*

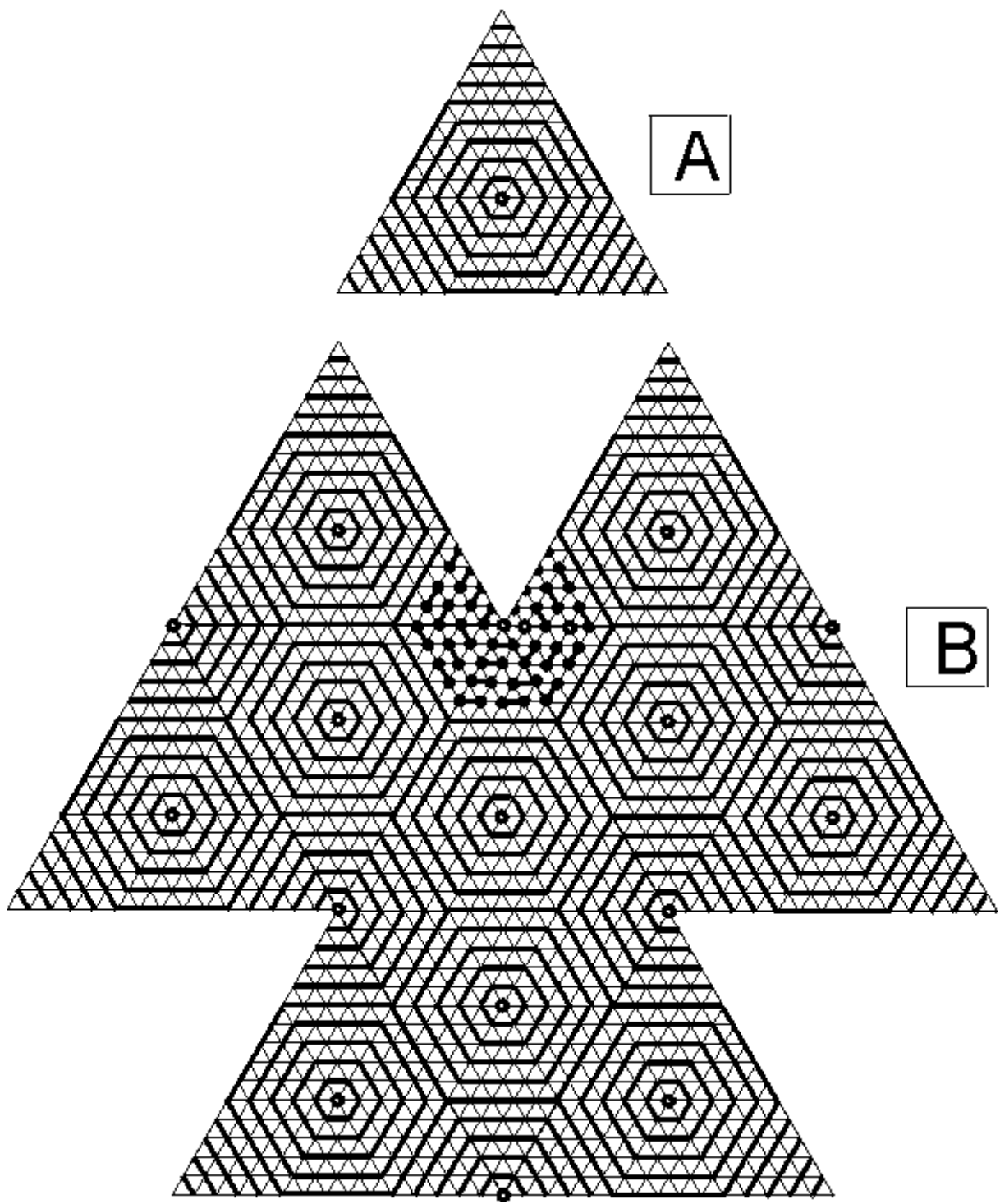

**Fig. 14**

*Configurations of phospholipids in triangles and icosahedrons. The linear and hexagonal arrays are represented schematically by solid lines and not by diglycerids as they should be.*

*A) This figure shows a triangle with $n_T=15$ that is filled with linear arrays in addition to a hexagonal structure with $n_H=5$. A prerequisite to apply the icosahedron model to monolayers is that the latter can be represented by 20 identical triangles. And each triangle in turn is defined by its phospholipid configuration. This configuration shows the 3-fold symmetry of a regular triangle.*

*B) The configuration of the PL formed by ten identical triangles is illustrated with only one of the two parts of Fig. 13. Note that there are structural vacancies in the 12 corners of the icosahedron. Six of these vacancies are given in this figure. The 12 vacancies are the only lattice points on the vesicles, surrounded by 5 instead of 6 other lattice points. The whole icosahedron has icosahedral symmetry as far as the phospholipid configuration is concerned.*

*Open ends of the arrays are not present and therefore it is postulated that this configuration represents the lowest energy for $n_T=15$.*

*Another aspect can also be distinguished by replacing the lines around one corner by molecules of diglycerids. In general: the corners appear to have structural vacancies that are present in alternating linear arrays with an odd number of lattice points.*

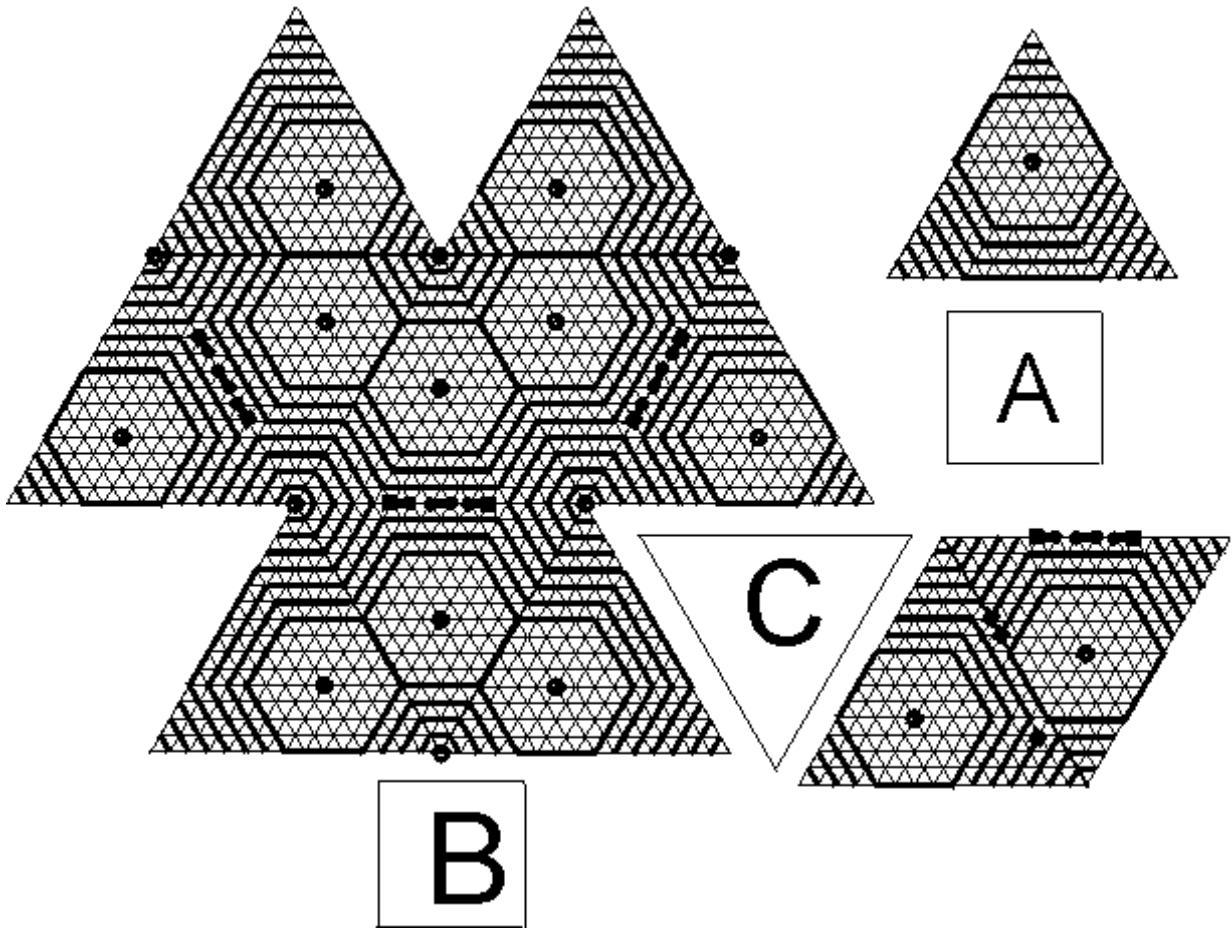

**Fig. 15**

*Like in Fig. 14 the arrays are represented by lines. Of the hexagonal structures only the hexagons with $n_H=4$ and their structural vacancies in the middle, are shown; the other hexagonal arrays formed by nesting are omitted.*

*A) Also this triangle has $n_T=15$. Because regular hexagons and linear arrays must be preferred the configuration in this triangle differs from the one in Fig. 14 A by the fact that the hexagon has $n_H=4$ and that this configuration cannot have 3-fold symmetry.*

*B) To make the difference with Fig. 14 as small as possible, in Fig A a configuration without open ends was constructed. In addition the linear arrays near the corners of every triangle were, if possible, arranged in the same way as shown in Fig. 14.*

*Despite these precautions linear arrays with open ends cannot be avoided when 20 triangles are assembled to an icosahedron. Therefore all configurations with $n_H=4$ have higher energy than the one of Fig. 14. Arrays with open ends are represented by 3 molecules of diglycerids. Chains of diglycerides at open ends, are represented by rectangles instead of by circles.*

*C) Extra open ends (here represented by squares) not yet present in the two-dimensional representation of Fig B, are created when two triangles are joined (cf. Fig 13) to form a three-dimensional icosahedron.*

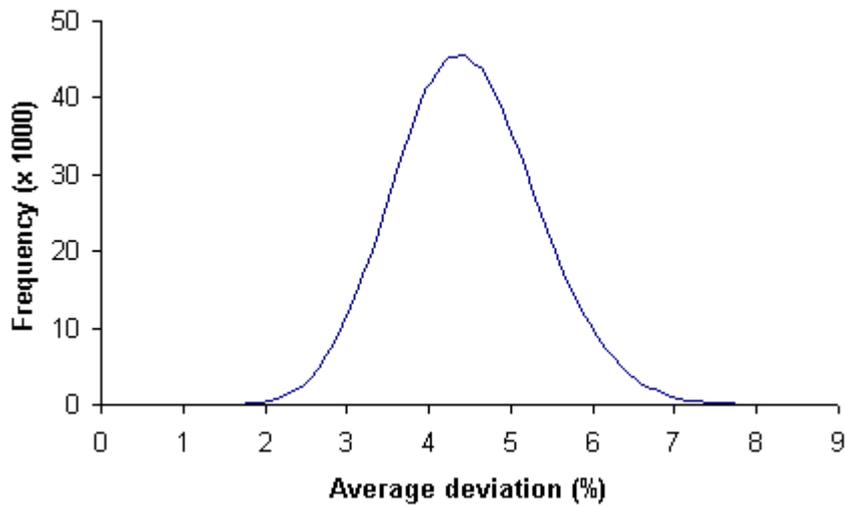

.

**Fig. 16**

*This figure represents the frequency distribution of average deviations of 12 random diameters and was constructed in the following way. In Table 2, theoretical diameters are given and in Table 3 these diameters are compared with the diameters of 12 in vivo vesicles and an average deviation of 1.8% calculated. In the same way the deviations of 12 random diameters from their theoretical counterparts (given in Table 2) were determined and this procedure was repeted one million times. The frequency of a certain average deviation was plotted against this deviation.*

**Table 1**

Comparison between the surface areas of respectively vesicles and one of their flat, triangular components.

The following vesicles are used: synaptic vesicles from cat [13]; synaptic vesicles from rat [8]; hormone vesicles from rat [11] and human hormone vesicles [14].

| $n_H$ | Surface area of triangles Series (b) | Surface area of triangles Series (c) | Diameter of vesicles D(V) | Effective surface area $\pi(D(V)-12)^2$ | Ratio between vesicles and triangles | |
|---|---|---|---|---|---|---|
| 12 | | 486 | 68.8 | 10136 | 20.8 | [13] |
| 8 | | 216 | 50.0 | 4536 | 21 | |
| 24 | | 1944 | 125.0 | 40115 | 20.6 | [8] |
| 32 | 1152 | | 97.5 | 22965 | 19.9 | |
| 48 | 2592 | | 147.5 | 57680 | 22.2 | [11] |
| 128 | 18432 | | 335.9 | 329588 | 17.9 | |

| | | | | | |
|---|---|---|---|---|---|
| 64 | 4608 | 181.6 | 90365 | 19.6 | |
| 128 | 18432 | 356.8 | 373494 | 20.2 | [14] |
| 96 | 31104 | 452.9 | 610703 | 19.6 | |
| | | | | | |

| | |
|---|---|
| Average | 20.2 |
| Standard deviation | 1.2 |

| Table 2 | Theoretical | diameters |
|---|---|---|
| $n_H$ | Calculated diameters of series b Equation (17) | Calculated diameters of series c Equation (18) |
| 8 | 33.4 | 49.1 |
| 12 | 44.1 | 67.6 |
| 16 | 54.8 | 86.1 |
| 24 | 76.2 | 123.2 |
| 32 | 97.6 | 160.3 |
| 48 | 140.4 | 234.4 |
| 64 | 183.3 | 308.7 |
| 96 | 268.9 | 457.0 |
| 128 | 354.5 | 605.3 |
| 192 | 525.8 | 902.0 |

**Table 3 Comparison between measured and calculated diameters**

| Source of diameter | Measured diameter | Calculated diameter Series b | Calculated diameter Series c | Deviation (%) |
|---|---|---|---|---|
| Synaptic vesicles cat [13] | 44.3 | 44.1 | | 0.4 |
| | 68.8 | | 67.6 | 1.7 |
| | 78.7 | 76.2 | | 3.1 |
| Synaptic vesicles rat [8] | 50.0 | | 49.1 | 1.8 |
| | 95.0 | 97.6 | | 2.7 |
| | 125.0 | | 123.2 | 1.4 |
| Hormone vesicles man [14] | 135.5 | 140.4 | | 3.6 |
| | 181.6 | 183.3 | | 0.9 |
| | 226.0 | | 234.4 | 3.7 |
| | 356.8 | 354.5 | | 0.6 |
| | 356.8 | 354.5 | | 0.6 |
| | 452.9 | | 457.0 | 0.9 |
| | | | Average = | 1.8 |